\documentstyle[12pt,epsfig]{article}
\newcommand{\beq}{\begin{equation}}
\newcommand{\eeq}{\end{equation}}
\begin{document}
\title{\bf SCALAR GLUEBALL: ANALYSIS OF THE
$(IJ^{PC}=00^{++})-$WAVE }
 \author{A.V.Anisovich, V.V.Anisovich, and A.V.Sarantsev
\\ St.Petersburg Nuclear Physics Institute\\
Gatchina, St.Petersburg 188350, Russia }

\date{}
\maketitle

\begin{abstract}
Basing on the previously performed
$K-$matrix analysis of experimental data, we investigate, in the
framework of the propagator matrix ($D$-matrix) technique, the
1100-1900 MeV mass region, where overlapping resonances
$f_0(1300)$, $f_0(1500)$,  $f_0\left(1530\pm{ 90\atop 250}\right)$,
and $f_0(1780)$ are located. Neccessary elements of the $D$-matrix
technique are developed. The $D$-matrix analysis confirms  previous
$K$-matrix results: in the region 900-1900 MeV five scalar/isoscalar
states are
located. Four of them are  members of the two $q\bar q$-nonets,
while one state is an extra
for the $q\bar q$ systematics, being a good
candidate for the lightest scalar glueball. The $D$-matrix analysis
shows that this extra state, a candidate for the
lightest scalar glueball, is dispersed, due to a mixing with $q\bar q$-
states, over three resonances:
$f_0(1300)$, $f_0(1500)$, and $f_0\left(1530\pm{ 90\atop 250}\right)$.
The broad resonance $f_0\left(1530\pm{ 90\atop 250}\right)$ is
a descendant of the lightest glueball carrying about $50\% $ of the
gluonium component, the rest of the gluonium is shared between
 $f_0(1300)$ and $f_0(1500)$.
\end{abstract}

\newpage

\section{Introduction}

At present time the mesons in the mass region 1000-2000 MeV are under
intensive experimental studies. The problem
these investigations face is the overlapping of resonances with
identical quantum numbers: the widths of resonances in this region
are comparatively large, of the same order as  mass
differences of neighbouring resonances \cite{pdg}. This leads to an
inevitable mixing of neighbouring resonances if they have common decay
channels.  Consideration of this mixing is needed for a restoration of
the initial-meson masses as well as for the determination of the quark
content of the mesons.
Definition of mixing is in particular important in the search for
exotic mesons, such as glueballs and hybrids. Glueball (or hybrid) may
be located in the vicinity of  $q\bar q$- mesons with identical
quantum numbers that raises their mixing and leads to the dispersing
of the glueball (or hybrid) component over several mesons.

This paper is devoted to the calculation  of scalar/isoscalar
resonance mixing, $IJ^{PC}=00^{++}$: experimental data prove
that in this wave there is a state which is superflous for $q\bar q-$
systematics \cite{km1900}, being possibly the lightest scalar glueball.
The idea that in the mass region 1500-1700 MeV the lightest
scalar glueball exists has been inspired by Lattice Gluodynamics
calculations \cite{ukqcd,ibm,close}. However, the quark degrees of
freedom are not taken into account in these calculations properly,
therefore they should be regarded as qualitative guide only:
the $q\bar q$-components can easily shift the glueball mass by
100-300 MeV.
Glueball/$q \bar q$-meson mixing was considered in refs.
\cite{genov,amsclo},
though without an implication of resonance decays into the mixing
machinary. The K-matrix analysis demonstrates \cite{km1900,km}
that the resonance decay processes may cause the mass shift by the
same value 100-300 MeV. The K-matrix amplitude, being unitary in the
 physical region and because of that treating the overlapping
resonances correctly, does not reproduce left-hand side singularities
of the amplitude which are related to the interaction of constituents.
This means a neccessity to complement the K-matrix consideration by an
analysis which restores the analyticity of the amplitude.

Here we perform an analysis of the $00^{++}$-wave in terms of
the propagator matrix (or D-matrix).
Corresponding technique, based on the
dispertion relation $N/D$-method, reconstructs the amplitude
with correct analytic properties. The detailed presentation of
propagator matrix method is made in Chapter 2: we consider the
mixing of overlapping resonances, the mass shifts caused by
mixing and decay processes, and the decomposition of the final
(i.e. physically observed) state in a series of initial states.
Before only the $00^{++}$-wave has been treated, however a
 generalization for other waves can be easily done, e.g.
 in the framework of the method given in ref.
\cite{deut,as}.

Our D-matrix analysis of $00^{++}$-wave is made in terms of
$q\bar q$-states, while K-matrix analysis of refs. \cite{km1900,km}
 used the hadron language.
When going from the hadronic K-matrix to the quark-antiquark
propator matrix,
the quark-hadron duality  problem should be considered. It can be
illustrated using standard quark potential model. In this model the
$q\bar q$-levels are defined by the potential which increases
infinitely at large $r$, $V(r)\sim \alpha r$ (see Fig. 1).  The
infinitely increasing potential of confinement leads to the infinite
set of stable $q\bar q$  levels: of course, this is a simplified
picture of what is seen in the experiment. Only the lowest $q\bar q$
levels are stable with respect to hadronic decays. The higher $q\bar q$
states, $00^{++}$ included, decay into hadronic channels: an excited
$(q\bar q)_a$-state produces  new $q\bar q$ pair,
then the quarks $(q\bar q)_a+q\bar q$ recombine into mesons which
leave the confinement trap, forming a continuus meson spectrum
at large distances. Conventional picture of the
decay is shown in Fig. 1b where the confinement interaction is depicted
as certain barrier: the interaction at $r<R_{confinement}$ creates the
$q\bar q$-spectrum, while at $r>R_{confinement}$ a continuus spectrum of
mesons takes place. In the experiment one can observe a continnus meson
spectrum, so to extract
information on $q\bar q$ spectra is a problem of \\
(i) transition from hadron language to that of quarks and gluons;\\
(ii) elimination of the influence of meson spectrum  on $q\bar q$-levels
at $r>R_{confinement}$.

Sections 3 and 4 are devoted to the consideration of this problem. In
Section 3 the necessary information is given concerning composite
systems. In Section 4 the rules of quark combinatorics are summarized
that allows to restore the quark content of a meson using the decays
$q\bar q-meson \to  two\;mesons$. The same decays may serve as a
signature of a glueball candidate. Section 5 is devoted to the
 calculation technique for
the transition diagrams which are responsible
for the mixing of levels created at $r<R_{confinement}$. In Section 6
the discussed technique is applied to the analysis of $00^{++}$ wave.
Using the quark language for the transition diagrams, we investigate
1100-1900 MeV mass region where four $00^{++}$ states are located.
Previously \cite{aaps}, the mass region 1100-1700 MeV was studied
with resonances
$f_0(1300)$, $f_0(1500)$, and $f_0\left(1530\pm{ 90\atop 250}\right)$
taken into consideration.
The present investigation confirms the result of ref. \cite{aaps}:
the pure (gluodynamic) glueball is dispersed over three above-mentioned
resonances, and the broad state $f_0\left(1530\pm{ 90\atop 250}\right)$
is a descendant of the glueball.
The resonance
$f_0 (1780)$ has a small glueball admixture.

\section{D-Matrix Technique}

Here the D-matrix technique is presented in detail. First, we consider
the propagator of non-stable particle (Section 2.1), then the
propagator matrix  for the two mixing resonances is constructed,
followed by the generalization for an arbitrary number of resonances
 (Sections 2.2 and 2.3). Examples of a complete resonance
overlapping are considered in Section 2.4: in this case one of the
resonances created as a result of a mixing accumulates the widths of
all other initial resonances.

 \subsection{Propagator of Non-stable Particle }

Propagator of non-stable particle (resonace propagator)
is given by the sum of diagrams shown in Fig. 2a, 2b, 2c,
and so on. Scalar resonance propagator is equal to

\beq
D(s)=(m^2-s-B(s))^{-1}\,.
\label{v1}
\eeq
Here $s=p^2$, where $p$ is the resonance four-momentum, and
 $m$ is the input-state mass;
$d(s)=(m^2-s)^{-1}$ is the propagator of an input state
related to the diagram of Fig. 2a, and $B(s)$ is the loop
diagram which describes the transition of  input state into two
particles.  For the decay of scalar resonance  into two scalar or
pseudoscalar particles, the loop diagram $B(s)$ may be written in
a form of the dispersion integral as follows:
\beq
B(s)=\int\limits_{(\mu_1+\mu_2)^2}^{\infty}\frac{ds'}{\pi}\;
\frac{g^2(s')\rho(s')}{s'-s-i0}\quad.
\label{v2}
\eeq
 $\mu_1$ and $\mu_2$ are meson masses in the loop diagram,
$g(s)$ is a vertex of the transition
{\it resonance} $\rightarrow$ {\it two
mesons}, and $\rho(s)$ is invariant phase space of the two-meson state:
\beq
\rho(s)=\frac{1}{16\pi s}
\sqrt{[s-(\mu_1+\mu_2)^2][s-(\mu_1-\mu_2)^2]}\quad.
\label{v3}
\eeq
Complex resonance mass squared is determined by
the propagator pole, $m^2-s-B(s)=0$. Below
standard notation for it is used:  $s=M^2\equiv M_R^2-i\Gamma
M_R$.  Near $s=M_R^2$, the real part of the loop diagram, $B(s)$,
can be expanded in a series over $(s-M_R^2)$:
\beq
D^{-1}(s)=(1+ReB'(M_R^2))(M_R^2-s)-ig^2(s)\rho(s)\quad,
\label{v4}
\eeq
$$M_R^2=m^2-ReB(M_R^2)\quad.$$
Here we take into account that
\beq ImB(s)=g^2(s)\rho(s)\,,\qquad
ReB(s)=P\int\limits_{(\mu_1+\mu_2)^2}^{\infty}\frac{ds'}{\pi}\;\frac{g^2(s')
\rho(s')}{s'-s}\quad.
\label{v5}
\eeq
The vertex $g(s)$
has only left singularities  of the scattering amplitude,
thus being non-singular function in the resonance region.
Smooth behaviour of $g(s)$ justifies standard Breit-Wigner approximation
for $D(s)$, with resonance width being equal to:
$M_R\Gamma=g^2(M_R^2)\rho(M_R^2)/(1+ReB'(M_R^2))$.

For the resonance decay into several channels, one should make the
following replacement in eq. (\ref{v1}):
\beq
B(s)\rightarrow\sum_{n}B^{(n)}(s)\,,
\label{v6}
\eeq
where index $n$ refers to different channels.

The loop diagram
$B^{(n)}(s)$ is determined by eq. (\ref{v2}), with a specification of
vertices and masses:
$g(s)\rightarrow g_n(s)$ and $\mu_i\rightarrow\mu_{ni}$.
The scattering amplitude in the channel  $n$ is determined as
\beq
A_n(s)=g^{(n)}(s)D(s)g^{(n)}(s)\,.
\label{v7}
\eeq
The quantity
$(g^{(n)}(M_R^2))^2\rho_n(M_R^2)/M_R=\Gamma_n$ defines  partial
width of the resonance.

\subsection{Mixing of Two Resonances    }

In the two-resonance case, the propagator of state 1
is determined by the diagrams of figs.
2a, 2b, 2c type (transitions {\it state 1}$\rightarrow $
{\it state 1}) and fig. 2d type (transitions
{\it state 1}$\rightarrow${\it state 2}).
The sum of these diagrams gives:
\beq
D_{11}(s)=\left(m_1^2-s-B_{11}(s)-\frac{B_{12}(s)B_{21}(s)}
{m_2^2-s-B_{22}(s)}\right)^{-1}\,.
\label{v8}
\eeq
Here $m_1$ and $m_2$ are  masses of  input states 1 and 2, while
loop diagram $B_{ab}(s)$
is determined by eq. (\ref{v2}), with the substitution $g^2(s)
\rightarrow g_a(s)g_b(s)$.
Let  the propagator matrix be introduced:
\beq
\hat D=\left|\begin{array}{ll}
D_{11}& D_{12}\\
D_{21}& D_{22}\end{array}\right|\,,
\label{v9}
\eeq
where the non-diagonal term $D_{ab}$ is the transiton propagator
$state\;b\to state\;a$.
The D-matrix is equal to:
\beq
\hat D=\frac{1}{(M_1^2-s)(M_2^2-s)-B_{12}B_{21}}
\left|\begin{array}{cc}
M_2^2-s,& B_{12}\\
B_{21},& M_1^2-s
\end{array}\right|\,.
\label{v10}
\eeq
The following notation is used here:
\beq
M_a^2=m_a^2-B_{aa}(s)\qquad\qquad a=1,2\;.
\label{v11}
\eeq
Zeros of the denominator of propagator matrix  provide the complex
masses of resonances created as a result of the initial-state mixing:
\beq
\Pi(s)=(M_1^2-s)(M_2^2-s)-B_{12}B_{21}=0\;.
\label{v12}
\eeq
Denote the complex masses of mixed states as $M_A$ ¨ $M_B$.

a) \underline{Model with constant $B_{ab}$}\\
As the first step, we consider a simple model: let us assume that loop
diagrams $B_{ab}(s)$ depend weakly  on $s$ in the region
 $s\sim M_A^2$ and $s\sim M_B^2$. Let $M_a^2$ and $B_{12}$
be  constants, then we have:
\beq
M_{A,B}^2=\frac 12 (M_1^2+M_2^2)\pm\sqrt{\frac 14
(M_1^2-M_2^2)^2+ B_{12}B_{21}}\quad.
\label{v13}
\eeq
Small widths of the resonances 1 and 2 affect small imaginary
part of $B_{12}$. In this case eq. (\ref{v13}) gives the standard
quantum mechanics result for splitting levels: a repulsion of mixed
levels.

The D-matrix can be written as a sum of pole terms,
 as follows:
\beq
\hat D=\left|\begin{array}{cc}
\frac{\cos^2\theta}{M_A^2-s}+\frac{\sin^2\theta}{M_B^2-s}&
\frac{-\cos\theta\sin\theta}{M_A^2-s}+\frac{\sin\theta\cos\theta}
{M_B^2-s}\\
~ & ~ \\
\frac{-\cos\theta\sin\theta}{M_A^2-s}+\frac{\sin\theta\cos\theta}
{M_B^2-s}&
\frac{\sin^2\theta}{M_A^2-s}+\frac{\cos^2\theta}{M_B^2-s}
\end{array}\right|\quad,
\label{v14}
\eeq
where
\beq
\cos^2\theta=\frac 12+\frac 12\frac{\frac 12(M_1^2-M_2^2)}
{\sqrt{\frac 14(M_1^2-M_2^2)^2+B_{12}B_{21}}}\quad.
\label{v15}
\eeq
The states $|A>$ and $|B>$ are superpositions
of initial states $|1>$ and $|2>$:
\beq
|A>=\cos\theta|1>-\sin\theta|2>\quad,
\label{v16}
\eeq
$$|B>=\sin\theta|1>+\cos\theta|2>\quad.$$

b) \underline{General case}
In general case,  when the
$s$-dependence of loop diagrams is not negligible and
$ImB_{ij}(s)$ is not small,
the states $|A>$ and $|B>$ can be presented as a superposition
of initial states as well. For this aim,
let us consider the propagator matrix
near $s=M_A^2$:
\beq
\hat D=\frac{1}{\Pi(s)}\left|\begin{array}{cc}
M_2^2(s)-s&B_{12}(s)\\
B_{21}(s)&M_1^2(s)-s
\end{array}\right|\,\simeq
\label{v17}
\eeq
$$\simeq\,\frac{-1}{\Pi'(M_A^2)(M_A^2-s)}\cdot\left|\begin{array}{cc}
M_2^2(M_A^2)-M_A^2& B_{12}(M_A^2)\\
B_{21}(M_A^2)&
M_1^2(M_A^2)-M_A^2\end{array}\right|.$$
Only singular (pole) terms are taken into account here.
In the right-hand side of eq. (\ref{v17}), the matrix determinant
is equal to zero due to eq. (\ref{v12}). Indeed:
\beq
[M_2^2(M_A^2)-M_A^2][M_1^2(M_A^2)-M_A^2]-B_{12}(M_A^2)B_{21}(M_A^2)=0\quad.
\label{v18}
\eeq
Eq. (\ref{v12}), $\Pi(M_A^2)=0$, allows us to introduce the complex
mixing angle:
  \beq
|A>=\cos\theta_A|1>-\sin\theta_A|2>\,.
\label{v19}
  \eeq
Then, the right-hand side of eq. (\ref{v17})
can be  re-written, using $\theta_A$, as follows:
 \beq
 \left[\hat D\right]_{s\sim
M_A^2}=\frac{N_A}{M_A^2-s}\left|\begin{array}{cc}
\cos^2\theta_A&-\cos\theta_A\sin\theta_A\\
-\sin\theta_A\cos\theta_A&\sin^2\theta_A\end{array}\right|\,,
\label{v20}
\eeq
where
\beq
N_A=\frac{1}{\Pi'(M_A^2)}[2M_A^2-M_1^2-M_2^2]\,,
\label{v21}
\eeq
$$\cos^2\theta_A=\frac{M_A^2-M_2^2}{2M_A^2-M_1^2-M_2^2}\,,\qquad
\sin^2\theta_A=\frac{M_A^2-M_1^2}{2M_A^2-M_1^2-M_2^2}\,.$$
Remind that the functions
 $M_1^2(s)$, $M_2^2(s)$ and
$B_{12}(s)$ in eq. (\ref{v21}) are fixed at $s=M_A^2$. In the case
under consideration, when the angle
$\theta_A$ is a complex magnitude, the quantaties $\cos^2\theta_A$ and
$\sin^2\theta_A$ do not relate to probabilities
$|1>$ and $|2>$ in $|A>$:
the factors $\sqrt{N_A}\cos\theta_A$ and $-\sqrt{N_A}\sin\theta_A$
are amplitudes for the transitions $|A>\rightarrow |1>$ and $|A>\rightarrow
|2>$. This means that corresponding probabilities
are equal to $|\cos\theta_A|^2$ and
$|\sin\theta_A|^2$.

To  decompose  the state $|B>$
over initial states $|1>$ and $|2>$, one should likewise use
 the propagator matrix near  $s=M_B^2$.
Representing $|B>$ as
\beq
|B>=\sin\theta_B|1>+\cos\theta_B|2>\,,
\label{v22}
\eeq
we have the following equation for $\hat D$ near
$s=M_B^2$:
 \beq
 \left[\hat D\right]_{s\sim M_B^2}=\frac{N_B}{M_B^2-s}\left|
\begin{array}{ll}\sin^2\theta_B &\cos\theta_B\sin\theta_B\\
\sin\theta_B\cos\theta_B &\cos^2\theta_B\end{array}\right|\,,
\label{v23}
\eeq
where
\beq
N_B=\frac{1}{\Pi'(M_B^2)}\left[2M_B^2-M_1^2-M_2^2\right]\,,
\label{v24}
\eeq
$$\cos^2\theta_B=\frac{M_B^2-M_1^2}{2M_B^2-M_1^2-M_2^2}\,,\qquad
\sin^2\theta_B=\frac{M_B^2-M_2^2}{2M_B^2-M_1^2-M_2^2}\,.$$
In eq. (\ref{v24}) the functions $M_1^2(s)$, $M_2^2(s)$ and $B_{12}(s)$
are fixed at $s=M_B^2$.

When  $B_{ab}(s)$ depends weakly on $s$ and this
$s$-dependence may be neglected, the angles $\theta_A$ and $\theta_B$
coinside.  But in general case the angles are different.  In this point
the propagator matrix formulae differ essentially from those
of standard quantum mechanics.

Another characteristic feature of the
discussed formulae is related to the mass shift of
mixed levels: in standard quantum mechanics, we have a
repulsion of levels associated with a conservation of the mean value,
$(E_1+E_2)/2$ ( see also eq.(\ref{v13})).
In the general case eq.(\ref{v12}) can
yield both the repulsion of mixed levels and their attraction.

The scattering amplitude in the one-channel case is determined
as
  \beq
A(s)=g_a(s)D_{ab}(s)g_b(s)\,.
\label{v25}
\eeq
For a multichannel case, one should introduce
\beq
B_{ab}(s)=\sum_{n}B_{ab}^{(n)}(s)\,,
\label{v26}
\eeq
where $B_{ab}^{(n)}$ is a loop diagram for the channel $n$
with the phase space $\rho_n$ and vertices
$g_a^{(n)}$, $g_b^{(n)}$. The partial scattering amplitude in
the channel $n$ is equal to
\beq
A_n(s)=g_a^{(n)}(s)D_{ab}(s)g_b^{(n)}(s)\,.
\label{v27}
\eeq

\subsection{ Propagator Matrix for an Arbitrary Number of
    Resonances }\

Matrix elements
$D_{ab}$ responsible for the transition of the initial state $b$
into the state $a$, being a sum of diagrams of fig. 2 type,
 satisfy a set of linear equations:
  \beq
D_{ab}=D_{ac}B_{cb}(s)(m^2_b-s)^{-1}+\delta_{ab}(m_b^2-s)^{-1}\;,
\label{v28}
\eeq
where $B_{ab}(s)$ is the loop diagram, and  $\delta_{ab}$ is
Kroneker unit matrix. Let us introduce the diagonal propagator
matrix $\hat d$ for
initial (non-mixed) states:
\beq
\hat d=\left |\begin{array}{cccc}
(m_1^2-s)^{-1} & 0 & 0 &\cdots\\
0 &(m_2^2-s)^{-1} &  0 &\cdots\\
0 &0 &(m_3^2-s)^{-1} &  \cdots\\
\vdots & \vdots & \vdots & \vdots
\end{array}\right |\;.
\label{v29}
\eeq
Then eq.(\ref{v28}) can be re-written in the matrix form,
\beq
\hat D= \hat D \hat B \hat d +\hat d\;,
\label{v30}
\eeq
therefor one has:
\beq
\hat D=\frac{I}{\hat d^{-1}-\hat B}\;.
\label{v31}
\eeq
The matrix $\hat d^{-1}$ is diagonal, so the inversed
propagator matrix
 $\hat D^{-1}=(\hat d^{-1}-\hat B)$ is written in the
following form:
\beq
\hat D^{-1}=\left |\begin{array}{cccc} M_1^2-s &
-B_{12}(s) & -B_{13}(s) &\cdots\\ -B_{21}(s) &M_2^2-s &  -B_{23}(s)
&\cdots\\ -B_{31}(s) &-B_{32}(s) &M_3^2-s &  \cdots\\ \vdots & \vdots &
\vdots & \vdots \end{array}\right |\;,
\label{v32}
\eeq
where $M^2_a$ is
determined by eq. (\ref{v11}).  Inverting the matrix, one has for
$D_{ab}(s)$:
\beq
D_{ab}(s)=\frac{(-1)^{a+b}\Pi_{ba}^{(N-1)}(s)}{\Pi^{(N)}(s)}\;.
\label{v33}
\eeq
Here  $\Pi^{(N)}(s)$ is the matrix determinant of $\hat D^{-1}$, and
$\Pi_{ba}^{(N-1)}(s)$ is the matrix supplement to the element
$[\hat D^{-1}]_{ba}$,
i.e. the determinant of the matrix $\hat D^{-1}$ with the excluded
 $b$th row and $a$th column.

Let us write down $\Pi^{(N)}(s)$     as an example
of the three-resonance case:
$$\Pi^{(3)}(s)=
(M_1^2-s)(M_2^2-s)(M_3^2-s)-(M_1^2-s)B_{23}
B_{32}-(M_2^2-s)B_{31}B_{13}\,-$$
\beq
-\,(M_3^2-s)B_{12}B_{21}-B_{12}B_{23}B_{31}-B_{13}B_{32}B_{21}\,.
\label{v34}
\eeq

Zeros of $\Pi^{(N)}(s)$ give poles of the propagator matrix,
which correspond to physical resonances created after
the  mixing of initial states. Denote the complex masses of resonances
(or zeros of $\Pi^{(N)}(s)$) as
\beq
s=M_A^2\,,\quad M_B^2\,,\quad
M_C^2\,, \ldots \label{v35}
\eeq
In the vicinity of the point
$s=M_A^2$, the pole terms provide the leading contribution. Neglecting
the next-to-leading terms in eq. (\ref{v30}), one has a system of
homogenious equations for $D_{ab}(s)$:
\beq
D_{ac}(s)\left ( \hat d^{-1}-\hat B \right)_{cb}=0
\label{v36}
\eeq
The solution of this system, found  with an accuracy up to
common normalization factor, has a factorized form:
\beq
 \left[\hat D^{(N)}\right]_{s\sim M_A^2}=\frac{N_A}{M_A^2-s}
\left|\begin{array}{llll}\alpha_1^2,&\alpha_1\alpha_2,&
\alpha_1\alpha_3, & \ldots\\
\alpha_2\alpha_1,&\alpha_2^2,&\alpha_2\alpha_3,& \ldots\\
\alpha_3\alpha_1,&\alpha_3\alpha_2,&\alpha_3^2,& \ldots\\
\ldots &  \ldots & \ldots & \ldots\end{array}\right|\,,
\label{v37}
\eeq
 $N_A$ is a normalization factor which is chosen to satisfy:
    \beq
\alpha_1^2+\alpha_2^2+\alpha_3^2+\ldots+\alpha_N^2=1\,.
\label{v38}
\eeq
The complex coupling $\alpha_a$ is a normalized transition
amplitude
\beq
\alpha_a(resonance\;A\rightarrow state\;a)\; ,
\label{v39}
\eeq
so the probability to find the state  $a$ in the resonance $A$ is
equal to:
 \beq
 W_a=|\alpha_a|^2\;.
\label{v40}
 \eeq
Analogous expansion of the propagator matrix can be done
in the vicinity of other poles:
  \beq
D_{ab}^{(N)}(s\sim M_B^2)=N_B\frac{\beta_a\beta_b}{M_B^2-s}\,,\qquad
D_{ab}^{(N)}(s\sim M_C^2)=N_C\frac{\gamma_a\gamma_b}{M_C^2-s}\,.
\label{v41}
 \eeq
The couplings satisfy the normalization conditions similar to
eq.(38):
 \beq
\beta_1^2+\beta_2^2+\ldots+\beta_N^2=1\,,\qquad
\gamma_1^2+\gamma_2^2+\ldots+\gamma_N^2=1\,.
\label{v42}
\eeq
But in general case,  the completeness condition is not fulfilled
for the inversed expansion:
 \beq
\alpha_a^2+\beta_a^2+\gamma_a^2+\ldots\neq 1\,.
\label{v43}
 \eeq
For the two-resonance case, this means that
 $\cos^2\Theta_A+ \sin^2\Theta_B\neq 1$.
Remind, however, that the right-hand side of eq.(\ref{v43}) is equal
to unity in the model with constant $B_{ab}$: see eqs.
(\ref{v13})-(\ref{v16}) which
correspond to the standard quntum mechnics formulae for mixed
levels.

\subsection{Complete Overlapping of Resonances: Effect of Accumulation
of Resonance Widths }

We consider here the examples which describe ideal situation with a
mixing of completely overlapping resonances. These examples demonstrate
in intact form the effect of width accumulation by one of resonances.

  )\underline{Complete overlapping of two resonances.}

Example 1:

For the simplicity sake, consider the
the weakly-s-dependent transition loop diagram, $B_{ab}$; this case
allows us to use eq. (\ref{v13}). We suppose
\beq
M_1^2=M_R^2-iM_R\Gamma_1\,,\qquad M_2^2=M_R^2-iM_R\Gamma_2\,,
\label{v44}
\eeq
that means
\beq
ReB_{12}(M_R^2)=P\int_{(\mu_1+\mu_2)^2}^{\infty}\frac{ds'}{\pi}
\frac{g_1(s')g_2(s')\rho(s')}{s'-M_R^2}=0\,.
\label{v45}
\eeq
Eq. (\ref{v45}) can be valid even
for positive $g_1$ and $g_2$ if the integral
over the region $s'<M_R^2$  compensates  the integral over $s'>M_R^2$.

In this case
\beq
B_{12}(M_R^2)=ig_1(M_R^2)g_2(M_R^2)\rho(M_R^2)=
iM_R\sqrt{\Gamma_1\Gamma_2}\,.
\label{v46}
\eeq
Substituting  (\ref{v44})-(\ref{v46}) into eq. (\ref{v13}), we have
$$
M_{A,B}^2=\frac 12(M_1^2+M_2^2)\pm\sqrt{\frac 14(-iM_R^2\Gamma_1+
iM_R^2\Gamma_2)^2+\left(iM_R\sqrt{\Gamma_1\Gamma_2}\right)^2}\,=
$$
\beq
=\,\left\{\begin{array}{l}
M_R^2-iM_R(\Gamma_1+\Gamma_2)\\
M_R^2\end{array}\right.
\label{v47}
\eeq
It is seen that, as a result of mixing, one state accumulates the
widths of both initial states, $\Gamma_A=\Gamma_1+\Gamma_2$, while
another one turns into a stable particle, $\Gamma_B=0$.

Example 2:

Consider one more example when
$Re\;M_1^2$ and $Re\;M_2^2$ are different but $M_1\Gamma_1=M_2\Gamma_2$,
namely:
\beq
 M_1^2=M_{R1}^2-iM\Gamma\,,\qquad M_2^2=M_{R2}^2-iM\Gamma\,.
\label{v48}
  \eeq
Then eq. (\ref{v13}) reads:
\beq
M_{A,B}^2=\frac 12(M_{R1}^2+M_{R2}^2)-iM\Gamma
\pm\sqrt{\frac 14(M_{R1}^2-M_{R2}^2)^2-M^2\Gamma^2}.
\label{v49}
\eeq
This equation allows one to see the dynamics of poles with an
increase of $\Gamma$. At $2M\Gamma\ll |M^2_{R1}-M^2_{R2}|$, that
corresponds to a suppressed mixing, one has two poles located near the
positions given by eq. (\ref{v48}).  With increasing $\Gamma$, the
poles move to each other along the real axis. At $2M\Gamma=
|M^2_{R1}-M^2_{R2}|$, the pole positions coinside with each other:
\beq
 M_{A,B}^2=\frac{1}{2}(M_{R1}^2+M_{R2}^2)-iM\Gamma
\label{v50}
\eeq
With further increase of
$\Gamma$, the poles move along the imaginary axis: as a result,
there are two poles, one above another. At $|M^2_{R1}-M^2_{R2}|\ll
2M\Gamma$, one state is almost stable while the width of another
resonance is close to $2\Gamma$.

b) \underline{Complete overlapping of three resonances}

Consider the equation
\beq
\Pi^{(3)}(s)=0
\label{v51}
\eeq
in the same approximation as in  Example 1.
Remind, $\Pi^{(3)}$
is determined by eq. (\ref{v34}). Correspondingly, we put
\beq
M_a^2=M_R^2-s-iM_R\Gamma_a=x-i\gamma_a \;,
\label{v52}
\eeq
$$Re B_{ab}(M_R^2)=0\,,\qquad (a\neq b)\,.$$
Here  $x=M_R^2-s$ and $M_R\Gamma_a=\gamma_a$.
Then, with $B_{ab}B_{ba}=i^2\gamma_a\gamma_b$ and
$B_{12}B_{23}B_{31}=i^3\gamma_1\gamma_2\gamma_3$, eq. (\ref{v34})  reads:
\beq
x^3+x^2(i\gamma_1+i\gamma_2+i\gamma_3)=0\,.
\label{v53}
\eeq
So, the propagator poles are at
\beq
M_A^2=M_R^2-iM_R(\Gamma_1+\Gamma_2+\Gamma_3),
\label{v54}
\eeq
$$M_B^2=M_C^2=M_R^2\,.$$
Resonance $A$ accumulates the widths of all initital resonances,
while the states $B$ and $C$ turn into being stable and degenerate.
Remark, the degenaracy goes off, when a weak $s-$dependence of loop
diagrams, $B_{ab}$, is taken into account.

\section{Composite systems}

Up to now, the resonance is treated as a mixture of input states,
and the loop diagrams $B_{ab}$ play crucial role in this mixing.
In addition, they take another function, for they determine the
compositeness of a particle.

\subsection{Entirely-composite particle}

Remind, following refs.\cite{deut} and
\cite{ammp}, the main features of the
N/D-method in a description of composite particles, like deuteron: this
gives us  an example of entirely-composite system. Entirely-composite
particle appeares as
a result of the interaction of  constituents and reveals itself as a
pole of the scattering amplitude of  constituents.  Correspnding
scattering amplitude for  multichannel case is
\beq
\frac{g^2_n
(s)}{1-B(s)}\,, \label{v55}
\eeq
 where $B(s)$ is a sum of loop diagrams
given by eq.  (\ref{v6}).  This amplitude differs from that considered
 in Section 2 by the replacement $(m^2-s)\to 1$:
  the absence of the input-particle propagator in the
scattering amplitude is a signature of an entirely composite-state.
 The mass of the composite state, $M$, is determined by the condition:
  \beq
 1-B(M^2)=0.
\label{v56}
\eeq
Suppose that the mass $M$ is located below all the thresholds, so we
deal with a stable composite particle. In the vicinity of
$s=M^2$, the scattering amplitude can be represented as
 a pole diagram of fig. 3d:
\beq
\frac{g^2_n (s)}{1-B(s)} \simeq \frac{g^2(s)}{B'(M^2)(M^2-s)}=
\frac{G^2_n (s)}{M^2-s}.
\label{v57}
\eeq
The vertex $G_n(s)$ describes the transition $composite\; particle\to
n-channel\;constitents$:
  \beq
 G_n(s)=\frac{g_n(s)}{\sqrt{B'(M^2)}}.
\label{v58}
 \eeq
The wave function of an entirely-composite particle is
the N-dimensional Fock column as follows:
\beq
\Psi=\left |\begin{array}{cccc}
\psi_1(s)\\
\psi_2(s)\\
\vdots     \\
\psi_N(s) \end{array}\right |\;,
\label{v59}
\eeq
where the partial wave function is determined as
\beq
\psi_n(s)=\frac{G_n(s)}{s-M^2}\;.
\label{v60}
\eeq
Definition of the partial wave function proceeds from
the calculation of the composite-particle electromagnetic form factor,
given by the sum of triange diagrams (fig. 3e), different
channels taken into account in the intermediate state (for detail, see
refs. \cite{deut,ammp}):
\beq
F(q^2)=\sum\limits_n\int\limits_{4\mu^2_n}^{\infty} \frac{ds\;ds'}{\pi^2}\;
\frac{G_n(s)\Delta_n (s,s',q^2)G_n(s')}{(s-M^2)(s'-M^2)}\;,
\label{v61}
\eeq
$$ \Delta_n (s,s',q^2)=\frac{(q^2-s-s')q^2}{16\lambda^{3/2}}\;
\Theta (-ss'q^2-\mu^2_n\lambda)\; ,$$
$$\lambda=s^2+s'^2+q^4-2ss'-2sq^2-2s'q^2\;.$$
Here, to avoid cumbersome expressions, we put the masses of
constituents in the intermediate state
equal to each other, $\mu_{1n}=\mu_{2n}\equiv \mu_n$.
The function $\Delta_n(s,s',q^2)$ in the limit  $q^2\to 0$
satisfies the equation:
\beq
\Delta_n(s,s',q^2\to 0)=\pi \rho_n(s)\delta (s-s')\;,
\label{v62}
\eeq
This leads to the following relation:
\beq
F(0)=\sum\limits_n\int\limits_{4\mu^2_n}^{\infty} \frac{ds}{\pi}\;
\frac{G_n(s)}{s-M^2}\;\rho_n (s)\;\frac{G_n(s)}{s-M^2}\;.
\label{v63}
\eeq
Substituting (\ref{v58}) into
right-hand side of eq.(\ref{v63}), one sees that $F(0)$ is equal to unity:
\beq
F(0)=\sum\limits_n\int\limits_{4\mu^2_n}^{\infty}
\frac{ds}{\pi}\; \rho_n (s)\;[\frac{G_n(s)}{s-M^2}]^2\;=1.
\label{v64}
\eeq

In terms of the  relative momenta of  constituents, eq.(\ref{v64})
reads:
\beq
1=\sum\limits_n\int \frac{d^3k_n}{(2\pi)^3}\;\psi^2_n(s)\;,
\label{v65}
\eeq
where $k^2_n=s/4-\mu^2_n$.
The quantity
\beq
W_n=\int \frac{d^3k_n}{(2\pi)^3}\;\psi^2_n(s)\;,
\label{v66}
\eeq
is the probability for the composite particle to be in the state $n$,
so
\beq
1=\sum\limits_n W_n\;.
\label{v67}
\eeq
This means that the composite particle is entirely built of the
constituents.

\subsection{Non-Entirely Composite Particle}

Now return to the case when the amplitude is determined
by a set of diagrams of fig. 2
but with a pole of the amplitude
located below all the thresholds. This means that we deal with a
stable particle. In this case the Fock
 column which describes  the wave function of the
composite particle contains  additional component, $\psi_0$, which
corresponds to the point-like input state with bare mass $m$:
\beq
\Psi=\left |\begin{array}{ccccc}
\psi_0  \\
\psi_1(s)\\
\psi_2(s)\\
\vdots     \\
\psi_N(s) \end{array}\right |\;,
\label{v68}
\eeq
Consideration of  $\psi_n(s)$ is made in a way similar
to previous case.
The scattering amplitude for  multichannel case is equal to
\beq
\frac{g^2_n (s)}{m^2-s-B(s)}\,.
\label{v69}
\eeq
Remind that $B(s)$ is  given by eq. (\ref{v6}). The
mass of the composite state, $M$, is determined by the following
condition:
\beq
m^2-M^2-B(M^2)=0.
\label{v70}
\eeq
Let us stress once again that $M$ is
 supposed to be below all the thresholds, so
this is a stable composite particle. In the vicinity of
$s=M^2$, the scattering amplitude can be represented as
\beq
\frac{g^2_n (s)}{m^2-s-B(s)} \simeq \frac{g^2(s)}{(1+B'(M^2))(M^2-s)}=
\frac{G^2_n (s)}{M^2-s}.
\label{v71}
\eeq
The vertex $G_n(s)$, responsible for the transition {\it composite
particle}$\to$ {\it n-channel constituents} is equal to:
   \beq
 G_n(s)=\frac{g_n(s)}{\sqrt{1+B'(M^2)}}\; .
\label{v72}
 \eeq
Partial wave function in the channel $n$ reads:
\beq
\psi_n(s)=\frac{G_n(s)}{s-M^2}\; .
\label{v73}
\eeq
As before, the definition of the partial wave
function proceeds from
the calculation of  electromagnetic form factor:
 \beq
F(q^2)=W_0+
\sum\limits_{n=1}^{N}
\int\limits_{4\mu^2_n}^{\infty} \frac{ds\;ds'}{\pi^2}\;
\frac{G_n(s)\Delta_n (s,s',q^2)G_n(s')}{(s-M^2)(s'-M^2)}\;,
\label{v74}
\eeq
 $ \Delta_n (s,s',q^2)$ is determined by eq. (\ref{v61}). Here, as compared
to eq.(\ref{v61}), an
additional constant term should be included into the form
factor, $W_0$, which corresponds to the interaction of  photon with a
point-like component (input particle with mass $m$): $W_0$ is the
probability for a particle to be in the input state. This leads to the
following normalization condition:
 \beq
F(0)=1=W_0+\sum\limits_{n=1}^N
\int\limits_{4\mu^2_n}^{\infty} \frac{ds}{\pi}\;
\frac{G_n(s)}{s-M^2}\;\rho_n (s)\;\frac{G_n(s)}{s-M^2}\;=
\label{v75}
\eeq
$$ =W_0 +\frac{B'(M^2)}{1+B'(M^2)}\;.$$

So, one has:
\beq
W_0 =\frac{1}{1+B'(M^2)}\;.
\label{v76}
\eeq
The probability for the considered particle to be in the state $n$
is equal to
\beq
W_n=\frac{B_{n}'(M^2)}{1+B'(M^2)}
=\int \frac{d^3k_n}{(2\pi)^3}\;\psi^2_n(s)\;.
\label{v77}
\eeq
The magnitude $\sum\limits_{n=1}^N W_n$
is the probability for a particle to be in the composite mode.

\subsection{Non-Stable Particles: Small and Large Distances in
Transition Diagrams}

Above we have discussed the  stable composite particle.
In the language of a potential, this corresponds to a level
inside potential well: see fig. 4a. Now consider  the case
of non-stable particle (resonance): corresponding shape of the
potential well and the level location is illustrated by fig.  4b.
The level is mostly created by the potential at $r<r_0$ (where $r_0$
corresponds
to the potential barrier maximum), while at $r\gg r_0$ the wave
function describes outgoing particles, which are the
resonance decay products.
 One may introduce the notions "inside the well" and
"outside the well" in the simplest way, using step-function
$\Theta (r_0-r)$, then the probaility for constituents to be bound is
determined by
\beq w_n =\int d^3 r\phi_n ^2(r)\Theta (r_0-r)\;.
\label{v78}
\eeq
Actually, we have introduced here the wave function
\beq
\phi^{in}_n(r)=\phi_n (r)\Theta (r_0-r)\;,
\label{v79}
\eeq
which is equal to $\phi_n(r)$ at $r<r_0$ and to zero at $r>r_0$.
Likewise, we can introduce the wave function for constituents at
$r>r_0$:
\beq
\phi^{out}_n(r)=\phi_n (r)\Theta (r-r_0)\;.
\label{v80}
\eeq
In the momentum representation, the  wave functions read:
\beq
 \psi^{in/out}_n(s)= \int d^3r\phi_n^{in/out}(r)e^{i\vec k\cdot\vec
r}\;,
\label{v81}
\eeq
where $\vec k=\frac{1}{2}(\vec k_1-\vec k_2)$ and
$\vec r=\vec r_1-\vec r_2$.

Presentation of the composite-particle wave function as a sum of
short- and
long-range pieces is the basic point in the analysis of $0^{++}$ wave.
The $q\bar q$-resonance being a composite state of quarks is
determined by the diagrams of the type shown in fig.
5a: quarks interact by gluon exchanges. The quark interaction at
$r<R_{confinement}$ results in the creation of bound states: in
our analysis we
approximate the small-$r$ amplitude as a set of pole terms (see
fig. 5b):
\beq
\sum\limits_a\frac{g_a^2}{m_a^2-s}\;.
\label{v82}
\eeq
 However, the sum of pole terms does not take into account the
interaction outside the confinement barrier, at $r>R_{confinement}$.
The quark loop with $r>R_{confinement}$ (fig. 5c) should be saturated
by a set of hadron loop diagrams (figs. 5d, 5e, and so on), in
consequence with quark-hadron duality.
Therefore, the inclusion of hadron loop diagrams into consideration
is a neccessary step in accounting for the large-$r$ quark interaction.

In terms of the $k$-representation, the separation of small-$r$ and
large-$r$ contributions occurs due to a choice of vertices $g_a(s)$:
suppression of the large-$s$ region means suppression of the
small-$r$ domain, and vice versa. Of course, the use of the step
function $\Theta (r-r_0)$ for a separation of the large-$r$
contribution is a simplification, which leads to
the violation of analytic properties of the transition loop
diagrams, $B(s)$, in the physical region:
 actually,  the small-$r$ region switch-off should
be done in a softer way. To keep the analyticity of $B(s)$'s, one
should care for analytic properties of  vertices, $g_a(s)$.

\section{The Couplings: Rules of $1/N$-Expansion and Quark Combinatoric
 Relations}

When analysing experimental data, we  approximate the amplitude at
$r>R_{confinement}$ by the set of pole terms, see eq. (\ref{v82}):
these poles
are determined by the quark-gluon interaction. Quark-gluon content of
the states related to poles reveals itself in  coupling constants
which govern the decay of these states into mesons, thus allowing to
find out from the experiment the quark-gluon composition of input
states.

In this Section, first,  the order of coupling constant
magnitude is evaluated within the rule of $1/N$ expansion \cite{thooft}
(Section 4.1).  Then, the decay constants are calculated (Section
4.2) using quark combinatoric rules. These latter were used rather long
ago for the calculation of secondaries in hadron-hadron collisions at
high energies \cite{anshe} and in the $J/\Psi$ decay \cite{JPsi}.
Calculation of the decay $ glueball \to two\;pseudoscalar\; mesons$
within the quark combinatorics was carried out in refs.
\cite{amsclo,gerst,vva2}. In the
same guideline, the decay couplings are calculated for the process
$ q\bar q -meson \to two\;pseudoscalar\; mesons$ in Section 4.3.

\subsection{Estimation of the Decay Couplings in the Framework of
$1/N$-Expansion Rules}

The rules of the $1/N$-expansion \cite{thooft} ($N=N_c=N_f$ where $N_c$
and $N_f$ are numbers of colours and light flavours) can be used as a
guide in the soft quark-gluon physics. Let us estimate, in terms of
$1/N$-expansion, the order of magnitude of the decay coupling
constants  $Glueball \to two\;mesons$ and $Glueball\to q\bar q$
denoted below as $g_{G\to mm}$ and $g_{~G\to q\bar q}$.

For this aim,  consider the gluon loop diagram in the glueball
propagator, fig. 6a, which is determined by the coupling $Glueball\to
two\; gluons$ denoted as $g_{G\to gg}$. This loop diagram is of the
order of unity if the glueball is a two-gluon composite system. From the
other side, this loop diagram is proportional to
$g^2_{G\to gg}N^2_c$ ($N^2_c$ is a number of colour states in a
 loop at large $N_c$), therefore
  \beq
 B(G\to gg\to G)\sim
g^2_{G\to gg}N^2_c \sim 1\;,
\label{v83}
 \eeq
 $$or\; \; g_{G\to gg}\sim 1/N_c \;. $$
The coupling $g_{\;G\to q\bar q}$ is determined by the diagram of
fig. 6b. Similar estimation leads to
\beq
g_{\;G \to q\bar q}\sim g_{\;G\to gg}g^2_{QCD}N_c\sim 1/N_c \;.
\label{v84}
\eeq
Here $g_{QCD}$ is gluon-quark coupling which is of the order of
$\sqrt{1/N_c}$, while the factor $N_c$ is related to the
number of  colour states in the triangle diagram.
Then the process shown in fig. 6c determines the coupling of the
$Glueball\to two\; mesons$ decay in the leading terms of
the $1/N_c$-expansion:
 \beq
  g_{G\to mm}^{L} \sim g_{\; G\to q\bar q}g^2_{m\to q\bar q}N_c
 \sim 1/N_c \;.
\label{v85}
 \eeq
Here we take into account that the coupling
$g_{m\to q\bar q}$ is of the order of $1/\sqrt{N_c}$: it is because
the quark loop diagram in the meson propagator $B(m\to q\bar q\to m)$
(see fig.6d) is of the order of unity:
\beq
B(m\to q\bar q \to m)\sim  g^2_{m\to q\bar q}N_c\sim 1  \;.
\label{v86}
\eeq

The diagram of fig. 6e provides the contribution which is of the
next-to-leading order of magnitude:
\beq
g^{NL}_{\; G\to mm}\sim g_{\; G\to gg}g^4_{QCD} g^2_{m\to q\bar q}
        N^2_c \sim 1/N_c^2.
\label{v87}
\eeq
Low-lying glueballs may decay into the following two-pseudoscalar
channels:
\beq
 n=\pi\pi,\;K\bar K,\;\eta\eta,\;\eta\eta',\;\eta'\eta'\;.
\label{v88}
\eeq
The ratios of couplings for the processes of fig. 6c can be
calculated in the framework of the quark combinatoric rules.
The important point here
 is the flavour symmetry violation in the  production of the
light quark pairs: $u\bar u$, $d\bar d$ and $s\bar s$. The production
probabilities refer as

\beq
u\bar u \; : \;d\bar d\; :\; s\bar s\;=\;1\;:\;1\;:\;\lambda,
\label{v89}
\eeq
with $\lambda=0.45\pm 0.05$ \cite{book,vva2}.
The couplings for the glueball decay into two pseudoscalar mesons
calculated within quark combinatorics are presented in Table 1 for
both leading and next-to-leading terms, $g^{L}_{G\to mm}$
and $g^{NL}_{G\to mm}$. Unknown dynamics of
the decay  is hidden into parameters $G^L$ and $G^{NL}$.
 The decay coupling to the channel $n$ is
equal to the sum of both contributions:
\beq
g^{L}_{G\to mm}(n)+g^{NL}_{G\to mm}(n)
\label{v90}
\eeq
The second term is suppressed, compared to the first one, by the
factor $1/N_c$: an experience with calculation of quark diagrams
gives us evidence that it leads to a suppression by a factor about
 $1/10$. The sums of coupling constants squared obey the
sum rules:
\beq
\sum\limits_n( g^{L}_{G \to
mm}(n))^2\;I(n)=\frac{1}{2}G_L^2 (2+\lambda )^2, \label{v91}
\eeq
$$ \sum\limits_n(g^{NL}_{G\to mm}(n))^2\;I(n)=\frac{1}{2}G^2_{NL}
(2+\lambda )^2  ,$$
where $I(n)$ is an identity factor for the particles produced
(see Table 1). These
sum rules reflect the fact that the decays under
consideration are resulted from the cutting of the two-loop
diagrams shown in figs. 6f and 6g, respectively: each quark loop
contains the factor $(2+\lambda)$, see eq. (\ref{v89}).

\subsection {Quark Combinatoric Calculations of the $q\bar q$-Meson
Decay Couplings}

Quark combinatoric rules can be applied to the $(q\bar q)_a$-meson decay
if we know the quark content of the meson. Let the
$(q\bar q)_a$-meson be the following mixture of non-strange and
strange quarks:
\beq
(q\bar q)_a=n\bar n\;cos\Phi\;+\;n\bar n\;sin\Phi \;,
\label{v92}
\eeq
 where $n\bar n=(u\bar u+d\bar d)/\sqrt 2$. Decay couplings of
this meson to two pseudoscalar meson channels, eq. (\ref{v88}), are
determined by the diagrams of fig. 7a and 7b types.

The process of fig. 7a provides the leading contribution, in terms of
 $1/N_c$-expansion:
\beq
g^L_{m(a)\to mm}\sim g_{m\to q\bar q}^3\;N_c \sim \frac{1}
{\sqrt{N_c}}.
\label{v93}
 \eeq
Similarly, one has for the process of fig. 7b:
 \beq
g^{NL}_{m(a)\to mm}\sim g_{m\to q\bar q}^3\;N_c^2
\; g^4_{m\to q\bar q}\sim \frac {1}{N_c\sqrt {N_c}}.
\label{v94}
\eeq
The couplings of the decays $m(a)\to
 \pi\pi,\;K\bar K,\;\eta\eta,\;\eta\eta',\;\eta'\eta'$ are presented
in Table 2 both for leading and next-to-leading terms: $g^L$ and
$g^{NL}$ are parameters in which the unknown dynamics of
the decay process is hidden.
The decay coupling to channel $n$ is a sum of the
leading and next-to-leading terms,
\beq
g^{L}_{m(a) \to mm}(n)+g^{NL}_{m(a)\to mm}(n)\;.
\label{v95}                                         
\eeq
These two terms are coupling constants in
the most general form, thus  giving all the varieties of decays. 
Examples with a special choice of coupling constants may be found in 
refs.  [16,18], where it was suggested that $g^{NL}_{m(a)\to mm}$ is 
not small as compared to $g^{L}_{m(a) \to mm}$.

 Let us emphasize a very important feature of the meson decay
couplings, $g_{m(a)\to mm}$: at $\Phi=\Phi_{Glueball}$ where
\beq
tan\Phi_{Glueball}=\sqrt{\frac{\lambda}{2}}\;,
\label{v96}
\eeq
the ratios of meson couplings coincide with those for the glueball
both for leading and next-to-leading terms. Namely,
\beq
[g^L_{m(a)\to mm}(n)]_{\Phi=\Phi_{Glueball}}\to g^L_{G\to mm}(n)\;,
\label{v97}
\eeq
$$[g^{NL}_{m(a)\to mm}(n)]_{\Phi=\Phi_{Glueball}}\to g^{NL}_{G\to
mm}(n)\;.  $$
It is due to the two-stage sructure of the glueball decay: on
the first stage, the $q\bar q$-pair is produced with the flavour content
given by eq. (\ref{v89}) that corresponds to $\Phi$ given by eq.
(\ref{v96}). The final
stage is a decay of this $q\bar q$-state. An important consequence of
the two-stage structure of the glueball decay reads:
basing on the ratio of the decay couplings to hadron channels only
it is impossible to find out
either we deal with the glueball or  $q\bar q$-state which has 
$n\bar n$/$s\bar s$ content close to that of eq. (\ref{v89}).

\section{Resonance Mixing in Terms of the Quark Transition Diagrams}

Quark--hadron duality means that one can analyse the $00^{++}$-wave
using quark or hadron languages. For our analysis this means the
choice of a language for the description of loop diagrams --- the
transitions may be described using either quark or hadron states. In
this Section the presentation of loop diagrams, $B_{ab}(s)$, is made in
terms of quark states. For the description of quark states, the light
cone variables are often used, so, as the first step, in Section 5.1
the loop diagram is written in these variables. As was mentioned above,
the quark model deals with wave functions in the $r$-representation; in
Section 5.2, $B_{ab}(s)$ is written using not vertices but wave
functions. Simple wave functions with one-parameter dependence are
presented in Section 5.3.

\subsection{Loop Diagram as a Function of Light Cone Variables}

Consider the loop diagram of eq.(\ref{v2}) using light cone
variables.  For this aim we write the phase space factor $\rho (s)$
in general form:  
\beq 
\rho (s)\to d\Phi(P,k_1,k_2)=  
\frac{1}{2} (2\pi)^4 \delta^{4}(P-k_1-k_2)
\frac{d^3k_1}{(2\pi )^3 2k_{10}}
\frac{d^3k_2}{(2\pi )^3 2k_{20}}
\label{v98}
\eeq
Rewriting $B(s)$ with light cone variables
can be done imposing $P_z\to \infty$. Then, denoting
\beq
k_{iz}/P_z\;=\;x_i\;,\;\;m^2+k^2_{\perp}=m^2_{\perp},
\label{v99}
\eeq
 one has for the right-hand side of eq.(\ref{v98}):
$$
\frac{1}{4(2\pi )^2} \frac{dx_1 dx_2}{x_1x_2} \delta (1-x_1-x_2)
d^2k_{1\perp}d^2k_{2\perp}
\delta (\vec P_{\perp}-\vec k_{1\perp}-\vec k_{2\perp})\cdot
$$
\beq
\cdot \delta (s+P^2_{\perp}
-\frac{m^2_{1\perp}}{x_1}-\frac{m^2_{2\perp}}{x_2}).
\label{v100}
\eeq
Putting $\vec P_{\perp}=0$, we rewrite $B(s)$ in terms of the light cone
variables:
 \beq
B(s)=
\frac{1}{4(2\pi )^2}\int \limits_{0}^{1}
\frac{dx_1 dx_2}{x_1x_2} \delta (1-x_1-x_2)
d^2k_{1\perp}d^2k_{2\perp}
\delta (\vec k_{1\perp}+\vec k_{2\perp})
\frac{g^2(s')}{s'-s-i0}\quad.
\label{v101}
\eeq
Here $s'=m^2_{1\perp}/x_1+m^2_{2\perp}/x_2$.

\subsection{Loop Diagram in the $r$-Representation}

To present the loop diagram as an integral in the coordinate space,
let us rewrite it in the c.m. $\vec k$-representation, where
$\vec k=(\vec k_1-\vec k_2)/2$ and
$\vec k_1+\vec k_2=0$. Then, the equation for $B_{ab}(s)$ reads:
 \beq
B_{ab}(s)=\int
\frac{d^3k}{(2\pi)^3}\;\frac{k_{10}+k_{20}}{2k_{10}k_{20}}
\;\frac{g_a(s')g_b(s')}{s'-s-i0}\quad,
\label{v102}
\eeq
$$
 s'=(k_{10}+k_{20})^2\;,\;k_{i0}=\sqrt{m_i^2+\vec k^2}\;.
$$
In terms of wave functions of the states $a$ and $b$, it is equal to:
\beq
B_{ab}(s)=\int
\frac{d^3k}{(2\pi)^3}\;\psi_a(s')
\;\frac{(s'-m^2_a)(s'-m^2_b)}{s'-s-i0}\psi_b(s')\;.
\label{v103}
\eeq
Here
 \beq
\psi_a(s)=
\sqrt{ \frac{k_{10}+k_{20}}{2k_{10}k_{20}} }
\;\frac{g_a(s)}{s'-m_a^2}\quad.
\label{v104}
\eeq
In the coordinate representation one has for $B_{ab}(s)$:
\beq
B_{ab}(s)=\int d^3rd^3r'\phi_a(r)\;v^{(ab)}(|\vec r-\vec r'|;s)
\phi_b(r')\;,
\label{v105}
\eeq
$$
v^{(ab)}(|\vec r|;s)= \int \frac{d^3k}{(2\pi)^3}\;
e^{i\vec r \cdot \vec k}
\;\frac{(s'-m^2_a)(s'-m^2_b)}{s'-s-i0}\;.
$$
The wave function in the coordinate representation is determined 
as usually:
\beq
\phi_a(r)=\int
 \frac{d^3k}{(2\pi)^3}\;e^{i\vec r\cdot \vec k}\psi_a(s)\;.
\label{v106}
\eeq
Constituent quark model makes it possible to estimate quantitatively
the $^3P_0q\bar q$ wave functions: making use of them,
 eqs. (\ref{v103}) and (\ref{v105}) allow us to calculate the
transition loop diagrams.

\subsection{Quark Transition Diagrams}

As was stressed above, when  the resonance
mixing has been analysed, two alternative languages can be used:
the hadron language or the quark one.
An advantage of the quark language is that the results of the quark
phenomenology may be used as a quantitative guide.

The quark loop diagram for the transition
$scalar\;state\; b \to$ $scalar$ $state$ $a$ is equal to:
\beq
B_{ab}(s)=\int\limits_{4m^2}^{\infty}\frac{ds'}{\pi}\;
\frac{g_a(s')\rho(s')g_b(s')}{s'-s-i0}\;
Tr[(\hat{k}+m)(\hat{P}-\hat{k}-m)]=
\label{v107}
\eeq
    $$
=\int\limits_{4m^2}^{\infty}\frac{ds'}{\pi}\;
\frac{g_a(s')\rho(s')g_b(s')}{s'-s-i0}\;2(s'-4m^2)\; ,
$$
where $m$ is  quark mass.
 In the trace calculations, we put $P^2=s'$ and
$k^2=m^2$:
there is no momentum conservation in the dispersion representation of
the loop diagram, but intermediate particles are on mass shell
 (for more detail, see refs. \cite{as,nato}).

In Section 3,
a receipt is given for the construction of composite-particle wave
function.
As a basis, one can use $B'_{aa}(M^2_a)$ where $M_a$ is the
composite-particle mass: this quantity is equal, up to  normalization
factor $\zeta$, to the wave function squared integrated over $d^3k$.
Applying this receipt to eq. (\ref{v107}), we get:
\beq
B'_{aa}(M^2_a)=\int
\frac{d^3k}{(2\pi)^3}[\frac{4g_a(s)}{s^{1/4}(s-M^2_a)}]^2\;
 (\vec S^{+} \cdot \vec k)
 (\vec S \cdot \vec k)\;\to
\label{v108}
\eeq
$$
\to \zeta \int
\frac{d^3k}{(2\pi)^3}\;\Psi_a^{+} (k)\Psi_a(k)\;.
$$
Here $\zeta =B'_{aa}(M^2_a)$ and
\beq
\Psi_a(k)=\frac{4G_a(s)}{s^{1/4}(s-M^2_a)}\;
 (\vec S \cdot \vec k)\;,
\label{v109}
\eeq
$$
G_a(s)=\frac {g_a(s)}{\sqrt {B'_{aa}(M^2_a)}}
$$
In eq.(\ref{v109}) we introduce the spin-1 operator $\vec S $.
 In standard representation it reads:
 \beq
 S_0=(0,0,1)\qquad\qquad
 S_{\pm 1}=(\frac{1}{\sqrt{2}}, \pm \frac{i}{\sqrt{2}},0)
\label{v110}
 \eeq
The wave function
\beq
 \Psi_a(k)= (\vec S \cdot \vec k)\psi_a(k)
\label{v111}
\eeq
can be rewritten
 in the r-representation as follows:
 \beq
 \Phi_a(r)= (\vec S \cdot \vec
 k)\phi_a(r) =\int \frac{d^3k}{(2\pi)^3}\;e^{-i\vec r\vec
k}\Psi_a(k)\;.
\label{v112}
  \eeq
 Here
\beq (\vec S \cdot \vec r)\phi_a(r)=i(\vec
 S\cdot \frac{\partial}{\partial \vec r})\varphi_a (r),
\label{v113}
 \eeq
 $$
\varphi_a(r)=\int
\frac{d^3k}{(2\pi)^3}\;e^{-i\vec r\vec k}\psi_a(k)\;.
$$
\underline{Examples}\\
Now we present simple examples
of the $N^3P_0q\bar q$-wave functions ( $N=1$, $2$), with a 
one parameter dependence. We suppose an  exponential decrease of wave 
functions at large $r$.  Then, the wave function of the $N=1$ state in 
the $r$-representation reads:  
\beq 
\Psi_{10^{++}}(r)=(\frac{2\alpha}{3})^{1/2} (\frac{\alpha}{\pi})^{3/4}
(\vec S \vec r) e^{-\frac{\alpha}{2}r^2}.
\label{v114}
\eeq
Parameter $\alpha$ is related to the mean radius squared $ <r^2>$ as
follows:
\beq
<r^2>_{1^3P_0}=\frac{5}{2\alpha}.
\label{v115}
\eeq
The same wave function in the $k^2$-representation reads:
\beq
\Psi_{10^{++}}(k^2)=\frac{4}{\sqrt {3}}(\frac{\pi}{\alpha})^{3/4}
\frac{i}{\sqrt {\alpha}}(\vec S \vec k)
e^{-\frac{k^2}{2\alpha}}.
\label{v116}
\eeq
The wave function of the $N=2$ state in the $r$- and $k$-representations
reads:
 \beq
\Psi_{20^{++}}(r)=(\frac{5\alpha}{3})^{1/2} (\frac{\alpha}{\pi})^{3/4}
(\vec S \vec r)(1-\frac{2}{5}\alpha r^2)
 e^{-\frac{\alpha}{2}r^2}\;,
\label{v117}
\eeq
$$
\Psi_{20^{++}}(k^2)=\frac{2\sqrt 2}{\sqrt {15}}
(\frac{\pi}{\alpha})^{3/4} \frac{i}{\sqrt {\alpha}}(\vec S \vec k)
(-5+2\frac{k^2}{\alpha})
e^{-\frac{k^2}{2\alpha}}\;.
$$
The mean radius squared of the $2^3P_0q\bar q$-state is equal to
\beq
<r^2>_{2^3P_0}=\frac{9}{2\alpha}\;.
\label{v118}
\eeq
Mean radii squared, eqs. (\ref{v115}) and (\ref{v118}), can be used
for fixing the wave functions of the $1 ^3P_0$ and $2 ^3P_0$ states.

\section{$D$-Matrix Amplitude and K-Matrix Representation}

In this Section we consider the connection between the D-matrix
amplitude and the amplitude in the K-matrix representation.
Usually the analysis of experimental data is performed in the
framework of the K-matrix amplitude: such an analysis has been carried 
out in refs. \cite{km1900,km}.  Performing a more complicated analysis, 
such as the D-matrix one, it is helpful to use the results of the 
K-matrix analysis at full scale.

In Section 6.1 the transition from D-matrix amplitude to the
K-matrix representation is considered
for the case of a separate resonance
decaying into a number channels. In Section 6.2 the
case of two overlapping resonances is investigated. The results of the
K-matrix analysis of the $00^{++}$ wave obtained in ref.\cite{km1900},
are quoted in Section 6.3. These results are used
in the next Section as a guideline for the construction of the D-matrix
amplitude.

\subsection{One-Resonance Amplitude}

The multichannel amplitude for a separate resonance is given
by eq.(\ref{v7}). The
transition from this amplitude to the K-matrix representation
is rather simple:
\beq
\sqrt{\rho_n(s)}A_{nn'}(s)\sqrt{\rho_{n'}(s)}
=\frac{g^{(n)}(s)g^{(n')}(s)\sqrt{\rho_n(s)\rho_{n'}(s)}}
{m^2-s-
\sum\limits_{n"}\left
(ReB^{(n")}(s)-i\rho_{n"}(s)g^{(n")2}(s)\right )}=
\frac{K_{nn'}(s)}{1-i\sum\limits_{n"} K_{n"n"}(s)},
\label{k1}
\eeq
where
\beq
K_{nn'}(s)=\frac{g^{(n)}(s)g^{(n')}(s)\sqrt{\rho_n(s)\rho_{n'}(s)}}
{m^2-s-\sum\limits_{n"} ReB^{(n")}(s)}\;.
\label{k2}
\eeq
Let us ephasize that the D-matrix amplitude gives factorized K-matrix
elements that means, for example, $K_{nn}K_{n'n'}=K_{nn'}K_{n'n}$.
This specific property results in a simplified K-matrix presentation
of the amplitude what is seen at eq. (\ref{k1}).
The K-matrix pole is defined by the equation:
\beq
m^2-s-ReB(s)=0\; ,\qquad\qquad B(s)\equiv\sum\limits_{n"} B^{(n")}(s).
\label{k3}
\eeq
Suppose that this equation is satisfied at some point $s=\mu^2$. Then
in the vicinity of this K-matrix pole the K-matrix elements are of
the form:
\beq
K_{nn'}(s)=\sqrt{\rho_n(s)}
\left [\frac{g^{(n)}(\mu^2)g^{(n')}(\mu^2)}
{(\mu^2-s)(1+ReB'(\mu^2))}+f_{nn'}\right ]
\sqrt{\rho_{n'}(s)}.
\label{k4}
\eeq
This is exactly the form used in refs.\cite{km1900,km} for  fitting
 experimental data. The pole position of the  K-matrix elements
is shifted, as compared to the propagator pole, in the magnitude
$ReB(\mu^2)$.  Position of the K-matrix pole also   differs 
significantly from the position of the amplitude pole, provided the 
value $ImB(s)$ is not suppressed.  This imaginary part should be 
saturated by meson-meson states, therefore their contribution can be 
considered as an additional "dressing" of  bound state by a cloud of 
real mesons.  Because of that we call K-matrix poles, which do not 
include this dressing, as poles of "bare" states \cite{km1900,km}.

Another important consequence of  eq.(\ref{k4}) is that the
K-matrix coupling constants are connected with the D-matrix couplings 
as follows:  
\beq 
g^{(n)}(K-matrix)=\frac{g^{(n)}(\mu^2)}{\sqrt{1+ReB'(\mu^2)}}\;.
\label{k5}
\eeq
Therefore, if the D-matrix couplings satisfy  quark combinatoric
relations, the same relations are valid for the K-matrix coupling
constants: the difference between $g^{(n)}(K-matrix)$ and
$g^{(n)}(\mu^2)$ lays only in the factor $(1+ReB'(s))^{1/2}$ which
is  common  for all channels.

\subsection{Two-Resonance/Two Channel Amplitude}

Now we consider  the case when two overlapping resonances decay into
two meson-meson channels. Being rather simple for the
investigation, this example contains all principal points of a
representation of the D-matrix amplitude in
the K-matrix form.

The D-matrix for the case under
discussion is given by eq.(\ref{v10}) while the
transition amplitude is written
in eq.(\ref{v27}). For the scattering
amplitude in the channel 1 one has:
\beq
\rho_1(s)A_{11}(s)=\frac{g_1^{(1)}(M^2_2-s)g^{(1)}_1
+g_1^{(1)}B_{12}g_2^{(1)} +g_2^{(1)}B_{12}g_1^{(1)}+
g_2^{(1)}(M^2_1-s)g^{(1)}_2}{(M_1^2-s)(M_2^2-s)-B_{12}B_{21}}=
\label{k6}
\eeq
$$
=\frac{ K_{11}+i(K_{12}K_{21}-K_{11}K_{22} }
{ 1-i(K_{11}+K_{22})+(K_{12}K_{21}-K_{11}K_{22}) }\; .
$$
Remind that  $M^2_a=m^2_a-B_{aa}(s)$.
The main question which should be answered here is: under what 
condition the quark combinatoric rules are valid for both D-matrix and 
K-matrix coupling constants? For this purpose,  consider in 
more detail the matrix elements $K_{11}$ and $K_{22}$.  The K-matrix 
form of eq.(\ref{k6}) gives us  the following expressions for $K_{11}$ 
and $K_{22}$:  
\beq 
K_{nn}=\frac{g_1^{(n)}(\mu_2^2-s)g_1^{(n)} 
+g_2^{(n)}(\mu_1^2-s)g_2^{(n)} +2ReB_{12}g_1^{(n)}g_2^{(n)}}
{(\mu_1^2(s)-s)(\mu^2_2(s)-s)-(ReB_{12})^2}\; ,
\label{k7}
\eeq
where
\beq
\mu^2_a(s)=m_a^2-ReB_{aa}(s)=m_a^2-\sum \limits_n ReB^{(n)}_{aa}(s)\;.
\label{k8}
\eeq
Below we show that the K-matrix couplings sustain the quark 
combinatoric rules  if
 $ReB_{12}$ is small . Indeed, in this case one has:
  \beq
K_{nn}\simeq \frac{(g_1^{(n)}(s))^2}{\mu_1^2(s)-s}+
\frac{(g_2^{(n)}(s))^2}{\mu_2^2(s)-s} \; .
\label{k9}
\eeq
An expansion of the right-hand side of eq. (\ref{k9}) near the K-matrix
poles gives:
  \beq
 K_{nn}\simeq
\frac{(g_1^{(n)}(K-matrix))^2}{\mu_1^2-s}+
\frac{(g_2^{(n)}(K-matrix))^2}{\mu_2^2(s)-s}+f_{nn}\; ,
\label{k10}
\eeq
where
\beq              
g^{(n)}_a(K-matrix)=\frac{g_a^{(n)}(\mu_a^2)}{\sqrt{1+ReB_{aa}'(\mu_a^2)}},
\label{k11}
\eeq
so the K-matrix couplings
$g^{(n)}_a(K-matrix)$ satisfy the same relations as
$g_a^{(n)}(\mu_a^2)$ because they differ by the factor
$\sqrt{1+ReB_{aa}'(\mu_a^2)}$ which is common for all channels.
The K-matrix elements in the form given by eq. (\ref{v10}) were used
in refs.\cite{km1900,km} for fitting experimental data.

One sees that using  quark combinatorics for the K-matrix
couplings is justified if $ReB_{12}$ is small or, more generally,
if the real parts of the non-diagonal loop transition diagrams are
small.
In the case of $q\bar q$-states, a suppression of the non-diagonal 
terms $ReB_{ab}(s)$ occurs because of orthogonality of $(q\bar q)_a$ 
and $(q\bar q)_b$ wave functions. For  mesons --- members of the same
nonet --- the suppression happens due to  cancellation of the
$n\bar n$- and $s\bar s$-loop diagrams (this cancellation is
precise, provided the flavour SU(3)-symmetry is not broken). For
mesons which are  members of different nonets, the suppression of
$ReB_{ab}(s)$ is the result of  orthogonality of radial wave
functions.

 The situation with  imaginary parts of  non-diagonal loop 
diagrams may be quite different: the imaginary parts are determined by 
decays into channels with real mesons, and therefore they strongly depend 
on the location of thresholds.  For example, two lowest 
K-matrix bare states, which are located below 1.2 GeV \cite{km1900,km}, 
despite of their being  members of the same nonet,
strongly mix due to large imaginary part of the loop transition
diagram.  The reason is that for the low-mass region 
only imaginary part of the non-strange quark loop diagram is properly
saturated by $\pi\pi$ mesons, while the imaginary part of $s\bar s$
loop diagram, which should be saturated by $K\bar K$, $\eta\eta$ and
$\eta\eta'$ mesons, is small because of the threshold suppression
of phase spaces, and it
is not able to cancel the $\pi\pi$ contribution.

Another situation with the D-matrix and K-matrix couplings
may happen when $q\bar q$ states mix with
a particle of another origin, for
instance, with a glueball. This
case is a subject of our further investigation.

\subsection{Results of K-Matrix Fit of $00^{++}$
Amplitude}

Analysis of scalar/isoscalar states in the mass
region below 1900~MeV was performed in the
framework of the K-matrix approach in ref. \cite{km1900} basing on the
following data set: GAMS data for
$\pi^-p\to\pi^0\pi^0n$ \cite{gams1}, $\eta\eta n$ \cite{gams2},
$\eta\eta'n$ \cite{gams3};
CERN-M\"unich data for $\pi^-p\to\pi^+\pi^-n$ \cite{cern};
Crystal Barrel data
for $p\bar p\to\pi^0\pi^0\pi^0$, $\pi^0\pi^0\eta$,
$\pi^0\eta\eta$ \cite{cbc1,cbc2}; BNL data for $\pi\pi\to K\overline K$
\cite{bnl}.
The $K$-matrix elements were used in the form similar to eq. (127):
\beq
K_{nn'}=\sum\limits_\alpha\frac{g^{(n)}_\alpha(K-matrix)
g^{(n')}_\alpha(K-matrix)}{\mu^2_\alpha-s} +f_{nn'}\;,
\label{k12}
\eeq
with five channels taken into account
 $n=\pi \pi$, $K\overline K$, $\eta\eta$, $\eta\eta'$, $\pi\pi\pi\pi$.
Couplings $g^{(n)}_\alpha(K-matrix)$
 are supposed to
obey  quark combinatoric
relations that allows  to reconstruct the $q\bar q$/glueball
content of bare states below 1900 MeV.
 The simultaneous fit of the two-meson spectra of
refs.\cite{gams1}-\cite{bnl} gives two solutions
which describe well the data sets:\\
{\bf Solution I.} Two bare states $f^{bare}_0(720\pm100)$ and
$f^{bare}_0(1260\pm30)$ are members of the $1^3P_0$ $q\bar q$-nonet, with
$f^{bare}_0(720)$ being $s\bar s$-rich state,
$\phi(720)=-69^o\pm12^o$ (the flavour wave function is defined as
$n\bar n Cos\phi+s\bar s Sin\phi$ ).
 The bare states $f^{bare}_0(1600\pm50)$ and $f^{bare}_0
(1810\pm30)$ are  members of the $2^3P_0$-nonet; $f^{bare}_0
(1600)$ is dominantly $n\bar n$-state: $\phi(1600)=-6^o\pm15^o$. The
state $f^{bare}_0(1235\pm50)$
is superfluous for $q\bar q$-classification, being a candidate
for the lightest glueball: its
couplings to two-\-meson channels obey quark combinatoric
relations for glueball.\\
Concluding, we have the following states in the solution I:
 \begin{equation}
 \begin{array}{cl}
 \mbox{Type of state} & \mbox{Flavour wave function}\\
  f_0^{bare}(720) \to 1^3P_0q\bar q  & 0.37n\bar n-0.93s\bar s\\
  f_0^{bare}(1260)\to 1^3P_0q\bar q  & 0.93n\bar n+0.37s\bar s\\
  f_0^{bare}(1600)\to 2^3P_0q\bar q  & 0.995n\bar n-0.10s\bar s\\
  f_0^{bare}(1810)\to 2^3P_0q\bar q  & 0.10n\bar n+0.995s\bar s\\
  f_0^{bare}(1235)\to  Glueball \to  & 0.91n\bar n+0.42s\bar s\ .
\end{array}
\end{equation}
For the glueball, "flavour wave function" refers the $q\bar q$
intermediate state in the decay process (see fig. 6c).

{\bf Solution II.} Basic nonet members are the same as in
solution I. The members of the $2^3P_0$-\-nonet are the following:
$f^{bare}_0(1235)$ and $f^{bare}_0(1810)$; both these states
have significant $s\bar s$-component: $\phi(1235)=42^o\pm10^o$ and
$\phi(1810)=-53^o\pm10^o$. The state $f^{bare}_0(1560\pm30)$ is
superfluous for the $q\bar q$-\-classification, being a good candidate for
the lightest glueball.

So, we have in the solution II the following set of states:
 \begin{equation}
 \begin{array}{cl}
 \mbox{Type of state} & \mbox{Flavour wave function}\\
  f_0^{bare}(720) \to 1^3P_0q\bar q  & 0.37n\bar n-0.93s\bar s\\
  f_0^{bare}(1260)\to 1^3P_0q\bar q  & 0.93n\bar n+0.37s\bar s\\
  f_0^{bare}(1235)\to 2^3P_0q\bar q  & 0.74n\bar n+0.67s\bar s\\
  f_0^{bare}(1810)\to 2^3P_0q\bar q  & 0.67n\bar n-0.74s\bar s\\
  f_0^{bare}(1560)\to  Glueball \to  & 0.91n\bar n+0.42s\bar s\ .
\end{array}
\end{equation}
For both solutions five scalar resonances
with nearly coinciding positions of the amplitude poles were found:
 \begin{equation}
 \begin{array}{cl}
\mbox{Resonance } & \mbox{ Position of  pole on the }\\
& \mbox{ complex-$M$ plane, in MeV units }\\

f_0(980) & 1015\pm15-i(43\pm8)\\
f_0(1300) & 1300\pm20-i(120\pm20)\\
f_0(1500) & 1499\pm8-i(65\pm10) \\
f_0(1750) & 1750\pm30-i(125\pm70)\\
f_0(1200/1600) & 1530^{+90}_{-250}-i(560\pm140)\ .
\end{array}
\end{equation}
Comparison of the positions of  bare states
and those of resonances shows that the observed states
are strong mixtures of  the bare states
and two/four-meson states into which these resonances decay.

Concluding this section, let us discuss the problem of applying
 quark combinatorics to the K-matrix decay couplings for
the bare states in the mass region 1200-1600 MeV,
$f_0^{bare}(1260),\; f_0^{bare}(1235)$ and $f_0^{bare}(1600)$: one of
them is supposed to be the glueball. This means that corresponding
non-diagonal loop diagrams for the transitions
$Glueball\to 1^3P_0q\bar q$ and
$Glueball\to 2^3P_0q\bar q$ are not small, and the arguments of
 Section 6.2 do not work. Nevertheless, eqs. (131) and (132)
show us that  quark combinaroric rules can be applied to these
resonances as well. The reason is that the discussed $q\bar q$-states
have approximately the same $n\bar n/s\bar s$ content as the
intermediate state in the glueball decay (the flavour wave function of
this intermediate state is written  in eqs. (131) and (132)).
In this case, it is obvious that quark combinatoric relations are
almost the same both for the K- and D-matrix amplitudes,
see eq. (\ref{k7}).

\section {D-Matrix Fit of the $00^{++}$ Amplitude}

In this Section we consider the mixing of the glueball state
with neighbouring $q\bar q$-states. We restrict our consideration
by the region 1100-1900 MeV: the calculation of the lowest state
$f_0(980)$ requires to take into account
the mechanism which saturates the imaginary part of the
quark-antiquark loop diagram by meson-meson states
in the precise form.
In the large-mass region, above the main
 meson-meson thresholds, the detailed
description of the saturation is not important.

\subsection{Parameters and Results}

For the calculation of the quark loop diagrams,  we
should fix the vertices $g_a(s)$. We parametrize the vertices
for the transition $state \;a \rightarrow n \bar n$
in the simple form:\\
\beq
\begin{array}{ll}
1^3P_0~q \bar q-state:&
g_1(s)= \gamma_1\; \sqrt[4] s \;\frac{k_1^2+\sigma_1}{k^2+\sigma_1};\\
~&~\\
2^3P_0~q \bar q-first\; state:&
g_2(s)= \gamma_2\; \sqrt[4] s \;\left [
\frac{k_2^2+\sigma_2}{k^2+\sigma_2} - d
\frac{k_2^2+\sigma_2}{k^2+\sigma_2+h} \right ];\\
~&\\
Glueball:&
g_3(s)= \gamma_3\; \sqrt[4] s \;\frac{k_3^2+\sigma_3}{k^2+\sigma_3};\\
~&\\
2^3P_0~q \bar q-second\; state:&
g_4(s)=g_2(s).
\end{array}
\label{7}
\eeq
Here $k^2=\frac{s}{4} -m^2$ and $k_a^2=\frac{m_a^2}{4} -m^2$
where $m$ is the constituent quark mass (fixed at
350 MeV for  non-strange quark
and at 500 MeV for strange one); $m_a$,
$\gamma _a$ and $\sigma_a$ are parameters. Factor $d$ is due to
 orthogonality of the $1^3P_0 q\bar q$ and $2^3P_0 q\bar q$
states: we put $Re\;B_{12}(s_0)=0$ at $\sqrt{s_0}=1.5$ GeV. (In the case
of the
$s$-dependent $B$-functions the orthogonality requirement for loop
transition diagrams cannot be fixed at all values of $s$).

The parameters $m_a$, $\sigma_a$, $h$ and $\gamma_a$ ($a=1,2,3$)
should be determined by  masses and widths of the physical
resonances.
However, the masses $m_a$
are approximately fixed by  the K-matrix fit of
ref. \cite{km1900}, where masses of the K-matrix poles,
$\mu_a^{bare}$, are defined: $\mu_a^2 \simeq m_a^2-
B_{aa}^2(\mu_a^2)$ (see also eqs. (131) and (132)). Let us stress that
$m_3$ is the mass of  pure gluonic glueball which
is a subject of Lattice QCD calculation.

Parameters which are found in our fit of the $00^{++}$ amplitude in
the mass region 1200-1900 MeV are given in Table 3 for
the solutions I and II.
Using these parameters, we calculate the couplings $\alpha_a$
which are introduced by eqs. (\ref{v37}) and (\ref{v38}):
\beq
    |f_0(1300)>\to \\
    |f_0(1500)>\to \\
    |f_0(1530)>\to \\
    |f_0(1780)>\to  \; .
\eeq
These couplings determine
 relative weight of the initial state $a$ in the physical resonance
$A$:
\beq
W_a(A)= |\alpha_a|^2
\label{10}
\eeq
The probabilities  $W_a$ are given in Table 3 together
with masses of  physical resonances, $M_A$, and  masses of
 input states, $m_a$.

\subsection{Glueball/$q \bar q$-State Mixing}

In order to analyze the dynamics of the glueball/$q \bar q$
mixing, we use the following method: in final formulae the
vertices are replaced in a way:
\beq
g_a(s) \rightarrow  \xi g_a(s),
\label{11}
\eeq
with a factor $\xi$ running in the interval $0 \leq \xi \leq 1$.
The case $\xi=0$ corresponds to switching off the mixing of input
states. Input states are stable in this case, and
corresponding poles of the amplitude are at $s_a=m_a^2$. Fig. 8
shows the pole position at $\xi=0$ for  solution I (fig. 8a)
and  solution II (fig. 8b). For the glueball state, $m_3$ is
the mass of a pure glueball, without $q \bar q$ degrees of
freedom. In  solution I the pure-glueball mass is equal to
\beq
m_{pure\;glueball}(Solution\;I)=1225\;MeV,
\eeq
that definitely disagrees with Lattice-Gluodynamics
calculations for the lightest glueball. In solution II
\beq
m_{pure\;glueball}(Solution\;II)=1633\;MeV.
\eeq
This value is in a good agreement with recent
Lattice-Gluodynamics results:\\
1570$\pm$85(stat)$\pm$100(syst) MeV \cite{ukqcd,close} and
1707$\pm$64 MeV \cite{ibm}.
With increasing $\xi$ the poles are shifted into  lower part
of the complex mass plane. Let us discuss in detail the
solution II which is consistent with Lattice result.

At $\xi \simeq 0.1-0.5$ the glueball state of solution II is
mixing mainly with $2^3P_0$ $q \bar q$-state, at $\xi \simeq
0.8-1.0$ the mixture with $1^3P_0$ $q \bar q$-state  becomes
significant. As a result, the descendant of the pure glueball
state has the mass $M=1450-i450$ MeV. Its gluonic content is
47\% (see Table 3). We should emphasize: the
definition of $W_a$  suggests that
$\sum_{A=1,2,3,4} W_{glueball}(A) \neq 1$ because of the
s-dependent $B_{ab}$
in the propagator matrix.

Hypothesis that the lightest scalar glueball is strongly mixed
with neighbouring $q\bar q$ states  was
discussed previously (see refs.
 \cite{amsclo}, \cite{Uspekhi}, and
 references therein).
However, the attempts to reproduce a quantitative picture
of the glueball/$q\bar q$-state mixing and the mass shifts
caused by this mixing cannot be
successful within standard quantum mechanics approach
that misses two phenomena:
\\ (i) Glueball/$q\bar q$-state mixing
described by the propagator matrix can give both a repulsion of the 
mixed levels, as in the standard quantum mechanics, and an attraction 
of them.  The latter effect may happen because the loop diagrams 
$B_{ab}$ are complex magnitudes, and the imaginary parts $ImB_{ab}$ are 
rather large in the region 1500 MeV.  
\\ (ii) Overlapping resonances 
yield a repulsion of the amplitude pole positions along imaginary-s 
axis.  In the case of full overlap of two resonances the width of one 
state tends to zero, while the width of the second state
tends to the sum of
the widths of initial states, $\Gamma_{first} \simeq 0$ and $\Gamma_{second}
\simeq \Gamma_1+ \Gamma_2$. For three overlapping resonances the widths of two
states tend to zero,
 while the width of the third state accumulates the widths
of all initial resonances, $\Gamma_{third} \simeq \Gamma_1+ \Gamma_2 +
\Gamma_3$.

Therefore, in the case of  nearly overlapping
resonances, what occurs  in the region near 1500 MeV, it
is inevitable to have one resonance with a large width. It is
also natural that it is the glueball descendant with large width:
the glueball mixes with the neighbouring $1^3P_0$ $q \bar q$ and
$2^3P_0$ $q \bar q$ states, which are both $n \bar n$ rich,
without suppression.

\section{Conclusion}

On the basis of N/D method we have developed
the matrix propagator approach (D-matrix method)
for the analysis of hadron spectra in the case of overlapping
resonances. This technique preserves the unitarity of
a scattering amplitude as well as its analyticity in the
whole complex-$s$ plane. Being rather similar to the
K-matrix technique for the case of overlapping $q\bar q$
resonances the D-matrix technique can give  different
results when a particle of another nature (e.g. glueball or hybrid)
interacts with a $q\bar q$ resonance. The case of the interaction
of the scalar glueball with neighbouring $q\bar q$ states in the region
1100-1800 MeV is considered on the language of quarks and gluons.
The content of the initial $q\bar q$ and glueball states in the
physical resonances was calculated. It appeared that
the lightest gluodynamic glueball is dispersed over  neighbouring
resonances mixing mainly to $1^3P_0$ $q \bar q$ and
$2^3P_0$ $q \bar q$ states. With this mixing the glueball
descendant transforms into  broad resonance,
$f_0(1530^{+90}_{-250})$. This resonance contains (40-50)$\%$
of the glueball component. Another part of the glueball
component is shared between comparatively narrow resonances,
$f_0(1300)$ and $f_0(1500)$ which are descendants of $1^3P_0$
$q \bar q$ and $2^3P_0$ $q \bar q$ states.

The important development for understanding of meson spectra
would be the direct application of the D-matrix approach
(instead of the K-matrix one) to the analysis of the meson
production amplitude. Such approach  will
give not only the correct position and mixing of the exotic states
with the $q\bar q$-mesons but also clarify the situation at
low energies where a
usual K-matrix approach starts to violate  analytical properties
of the scattering amplitude.

\section{Acknowledgement}

We thank D.V.Bugg, F.E.Close, L.G.Dakhno, S.S.Gershtein, A.K.Likhoded,
L.Montanet, and Yu.D.Prokoshkin for heplful discussions and remarks.
We are indebted to H.Koch for kind invitation to participate at
LEAP'96-Conference (August 27-31, 1996, Dinkelsbuehl, Germany) where
prelimenary results of the analysis were presented.
This work was supported by  Russian Fundamental Investigation Fund,
grant 96-02-17934.

\newpage
\begin{center}
Table 1\\
Coupling constants given by quark combinatorics for a glueball
decaying into two pseudoscalar mesons in the leading
and  next-to-leading terms of the $1/N$ expansion.
$\Phi$ is the mixing angle for $n\bar n$ and $s\bar s$ states, and
$\Theta$ is the mixing angle for $\eta -\eta'$ mesons:
$\eta=n\bar n \cos\Theta-s\bar s \sin\Theta$ and
$\eta'=n\bar n \sin\Theta+s\bar s \cos\Theta$.
\vskip 0.5cm
\begin{tabular}{|c|c|c|c|}
\hline
~      &     ~                    &  ~                     &~           \\
~      & The glueball decay       &The decay couplings     &Identity   \\
~      & couplings in the         &in the next-to-         &   factor in  \\
Channel& leading terms of $1/N$   &leading terms of $1/N$ &phase space \\
~      & expansion (Fig. 1e)      &expansion (Fig. 1f)     &~   \\
~      &     ~                    &  ~                     &~           \\
\hline
~ & ~ & ~ & ~ \\
$\pi^0\pi^0$ &
$G^L$ & 0 & 1/2  \\
~ & ~ & ~ & ~ \\
$\pi^+\pi^-$ &
$G^L$ & 0 & 1  \\
~ & ~ & ~ & ~ \\
$K^+K^-$ & $\lambda G^L$ & 0 & 1 \\
~ & ~ & ~ & ~ \\
$K^0K^0$ & $\lambda G^L $ & 0 & 1 \\
~ & ~ & ~ & ~ \\
$\eta\eta$ &
$G^L\left (\cos^2\Theta+ \lambda\sin^2\Theta\right )$
&$G^{NL}(\cos\Theta-\sqrt{\frac{\lambda}{2}}\sin\Theta )^2$ &
1/2 \\
~ & ~ & ~ & ~ \\
$\eta\eta'$ &
$G^L (1-\lambda)\sin\Theta\;\cos\Theta$
&$G^{NL}(\cos\Theta-\sqrt{\frac{\lambda}{2}}\sin\Theta)\times$ &1 \\
~&~&$(\sin\Theta+\sqrt{\frac{\lambda}{2}}\cos\Theta)$ & ~\\
~ & ~ & ~ & ~ \\
$\eta'\eta'$ &
$G^L\left(\sin^2\Theta+\sqrt{\lambda}\;\cos^2\Theta\right)$
&$G^{NL}\left(\sin\Theta+\sqrt{\frac{\lambda}{2}}\cos\Theta \right)^2$ &
1/2 \\
~ & ~ & ~ & ~ \\
\hline
\end{tabular}
\end{center}

\newpage
\begin{center}
Table 2\\
Coupling constants given by quark combinatorics for a $(q\bar q)_a$-meson
decaying into two pseudoscalar mesons;
here $(q\bar q)_a=n\bar n \cos\Phi+s\bar s \sin\Phi$.
\vskip 0.5cm
\begin{tabular}{|c|c|c|}
\hline
~      &     ~                        &  ~                             \\
~      & The $(q\bar q)_a$-meson decay&The $(q\bar q)_a$-meson decay   \\
~      & couplings in the             &couplings in the next- \\
Channel& leading terms of $1/N_c$     &to-leading terms of $1/N_c$  \\
~      & expansion (Fig. 7a)          &expansion (Fig. 7b)    \\
~      &     ~                        &  ~                       \\
\hline
~ & ~ & ~  \\
$\pi^0\pi^0$ &
$g^L\;\cos\Phi/\sqrt{2}$& 0  \\
~ & ~ & ~  \\
$\pi^+\pi^-$ & $g^L\;\cos\Phi/\sqrt{2}$ & 0 \\
~ & ~ & ~ \\
$K^+K^-$ & $g^L (\sqrt 2\sin\Phi+\sqrt \lambda\cos\Phi)/\sqrt 8 $ & 0 \\
~ & ~ & ~ \\
$K^0K^0$ & $g^L (\sqrt 2\sin\Phi+\sqrt \lambda\cos\Phi)/\sqrt 8 $ & 0 \\
~ & ~ & ~ \\
$\eta\eta$ &
$g^L\left (\cos^2\Theta\;\cos\Phi/\sqrt 2+\right .$\hfill
&$2g^{NL}(\cos\Theta-\sqrt{\frac{\lambda}{2}}\sin\Theta )$\hfill\\
~ &\hfill$\left . \sqrt{\lambda}\;\sin\Phi\;\sin^2\Theta\right )$ &
\hfill$(\cos\Phi\cos\Theta-\sin\Phi\sin\Theta)$ \\
~ & ~ & ~  \\
$\eta\eta'$ &
$g^L\sin\Theta\;\cos\Theta\left(\cos\Phi/\sqrt 2-\right .$\hfill
&$\frac 12g^{NL}\left [(\cos\Theta-\sqrt{\frac{\lambda}{2}}\sin\Theta)
\right .$\\
~& \hfill $\left .\sqrt{\lambda}\;\sin\Phi\right ) $ &
\hfill$(\cos\Phi\sin\Theta+\sin\Phi\cos\Theta)$\\
~&~&$+(\sin\Theta+\sqrt{\frac{\lambda}{2}}\cos\Theta)$\hfill\\
~&~&\hfill$\left .(\cos\Phi\sin\Theta-\sin\Phi\cos\Theta)\right ]$\\
~ & ~ & ~  \\
$\eta'\eta'$ &
$g^L\left(\sin^2\Theta\;\cos\Phi/\sqrt 2+\right .$\hfill
&$2g^{NL} (\sin\Theta+\sqrt{\frac{\lambda}{2}}\cos\Theta)$\hfill\\
~&\hfill $\left .\sqrt{\lambda}\;\sin\Phi\;\cos^2\Theta\right)$ &
\hfill$(\cos\Phi\cos\Theta+\sin\Phi\sin\Theta )$ \\
~ & ~ & ~  \\
\hline
\end{tabular}
\end{center}
\newpage

\begin{center}
Table 3\\
\vskip 0.3cm
Masses of the initial states,  coupling constants
and $q\bar q$/glueball content of  physical states.
\vskip 0.3cm
\begin{tabular}{|c|c|c|c|c|}
\hline
\multicolumn{5}{|c|}{Solution I}  \\
\hline
~ & $1^3P_0$ & $2^3P_0$ & ~ & $2^3P_0$ \\
Initial state & $n\bar n$-rich & $n\bar n$-rich &
Glueball & $s\bar s$-rich\\
~ &$\phi_1=18^\circ$ &$\phi_2=-6^\circ$  & $\phi_3=25^\circ$ &
$\phi_4=84^\circ$\\
\hline
$m_i$ (GeV) & 1.457 & 1.536 & 1.230 & 1.750 \\
$\gamma_i$ (GeV$^{3/4}$) & 0.708 & 1.471 & 0.453 & 1.471 \\
$\sigma_i$ (GeV$^2$) & 0.075 & 0.225 & 0.375 & 0.225 \\
\hline
$W[f_0(1300)]$ & 32\% & 12\%  & 55\% & 1\%\\
$1.300-i0.115$ (GeV) & ~ &~ & ~& \\
$W[f_0(1500)]$ & 25\% & 70\%  & 3\% & 2\%\\
$1.500-i0.065$ (GeV) & ~ &~ & ~& \\
$W[f_0(1530)]$ & 44\% & 24\%  & 27\% & 4\%\\
$1.450-i0.450$ (GeV) & ~ &~ & ~& \\
$W[f_0(1780)]$ & 1\% &  1\%   & --    & 98\%\\
$1.780-i0.085$ (GeV) & ~ &~ & ~& \\
\hline
\multicolumn{5}{|c|}{$h=0.25$ GeV$^2$,~~~$d=1.01$}  \\
\hline
\hline
\multicolumn{5}{|c|}{Solution II}  \\
\hline
~ & $1^3P_0$ & $2^3P_0$ & ~ & $2^3P_0$ \\
Initial state & $n\bar n$-rich & $n\bar n$-rich &
Glueball & $s\bar s$-rich\\
~ &$\phi_1=18^\circ$ &$\phi_2=35^\circ$  & $\phi_3=25^\circ$ &
$\phi_4=-55^\circ$\\
\hline
$m_i$ (GeV)& 1.107 & 1.566 & 1.633 & 1.702 \\
$\gamma_i$ (GeV$^{3/4}$)& 0.512 & 0.994 & 0.446 & 0.994 \\
$\sigma_i$ (GeV$^2$) & 0.175 & 0.275 & 0.375 & 0.275 \\
\hline
$W[f_0(1300)]$ & 35\% & 26\%  & 38\% & 0.4\%\\
$1.300-i0.115$ (GeV) & ~ &~ & ~& \\
$W[f_0(1500)]$ & 1\% & 64\%  & 35\%  & 0.4\%\\
$1.500-i0.065$ (GeV) & ~ &~ & ~& \\
$W[f_0(1530)]$ & 12\% & 41\%  & 47\% & 0.3\%\\
$1.450-i0.450$ (GeV) & ~ &~ & ~& \\
$W[f_0(1780)]$ & 0.1\%  &  0.2\%   & 0.2\%   & 99.5\%\\
$1.750-i0.100$ (GeV) & ~ &~ & ~& \\
\hline
\multicolumn{5}{|c|}{$h=0.625$ GeV$^2$,~~~$d=1.16$}  \\
\hline
\end{tabular}
\end{center}
\newpage
\begin{center}
\epsfig{file=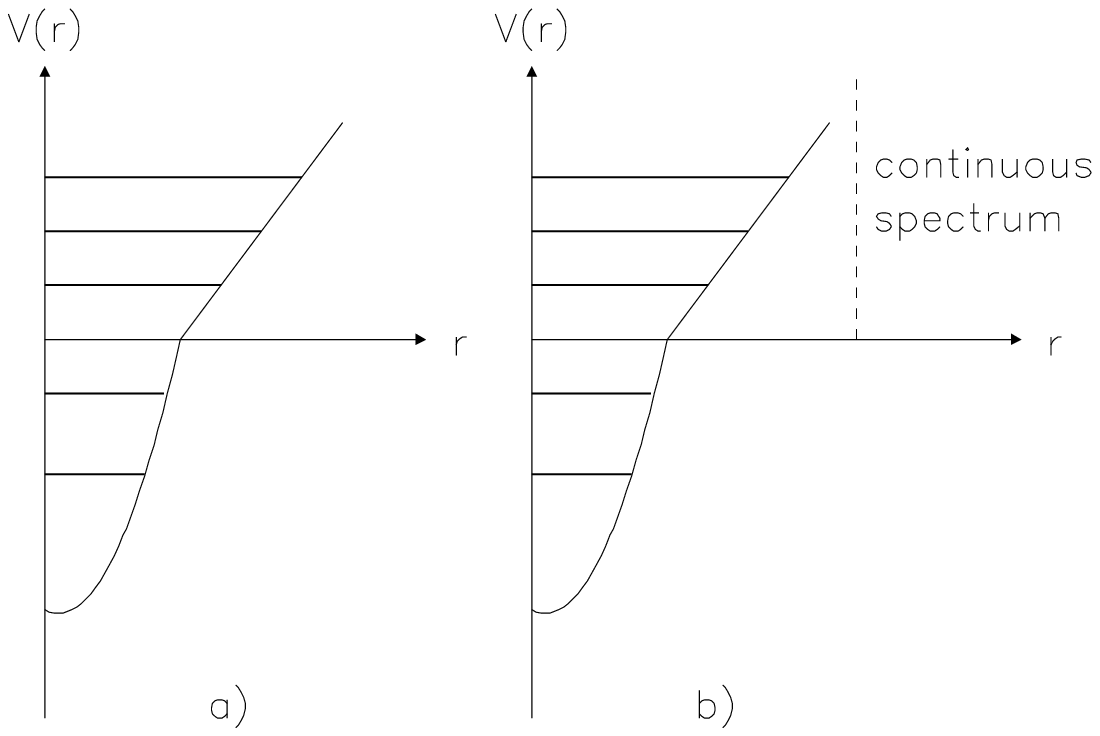,width=14cm}\\
Fig. 1. c) Meson states in the infinitely increasing quark model
potential b) Conventional picture for the transition $q\bar q-states$
$\to$ {\it meson states}.
\end{center}
\newpage
\begin{center}
\epsfig{file=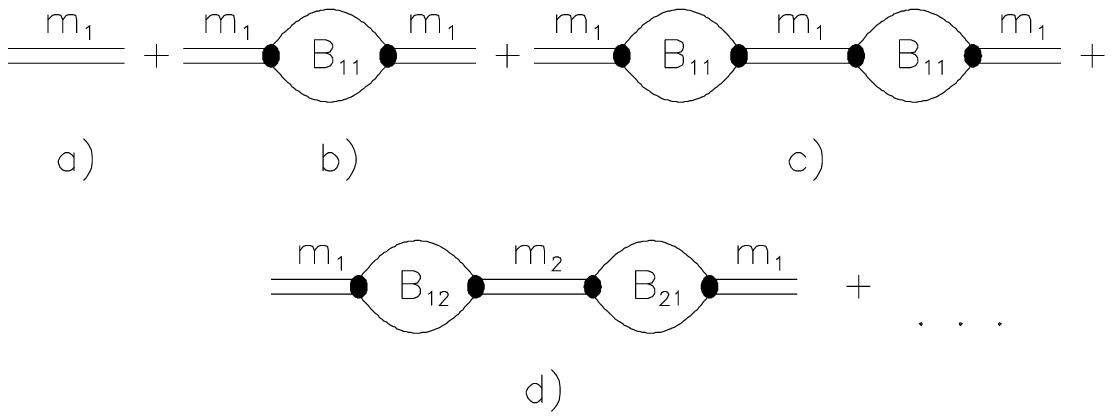,width=14cm}\\
Fig. 2. Diagrams responsible for the decay/mixing process.
\end{center}
\newpage
\begin{center}
\epsfig{file=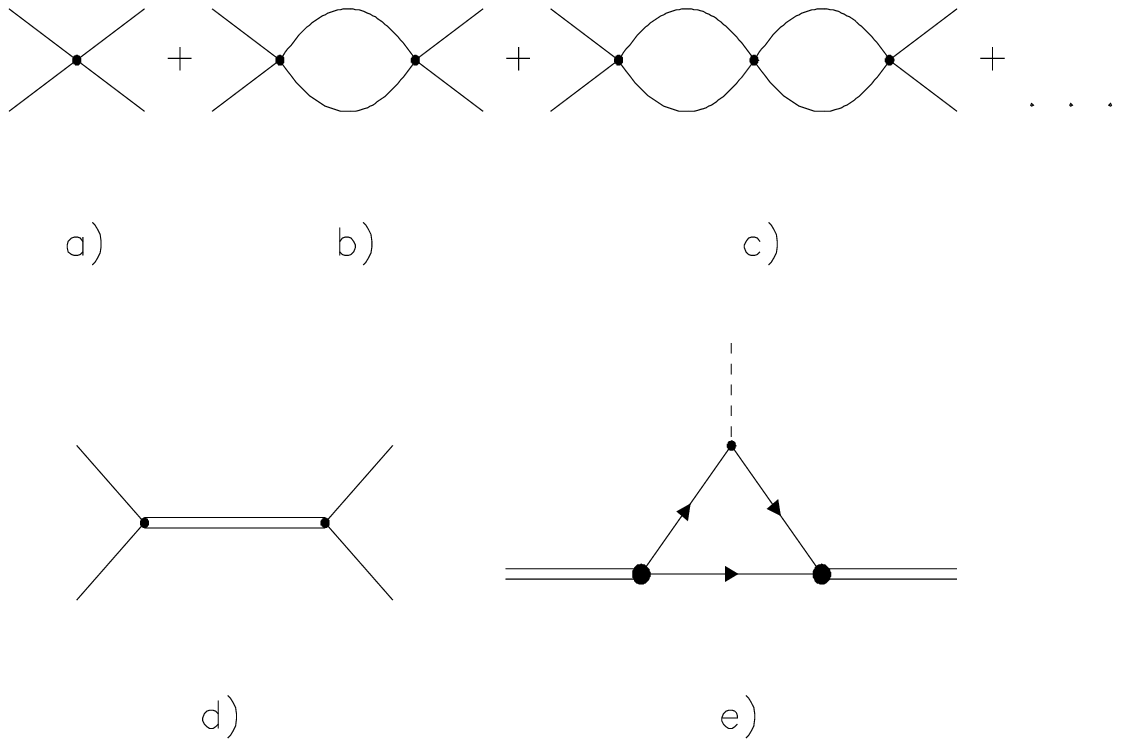,width=14cm}\\
Diagrams for entirely-composite particle.
Fig. 3.
\end{center}
\newpage
\begin{center}
\epsfig{file=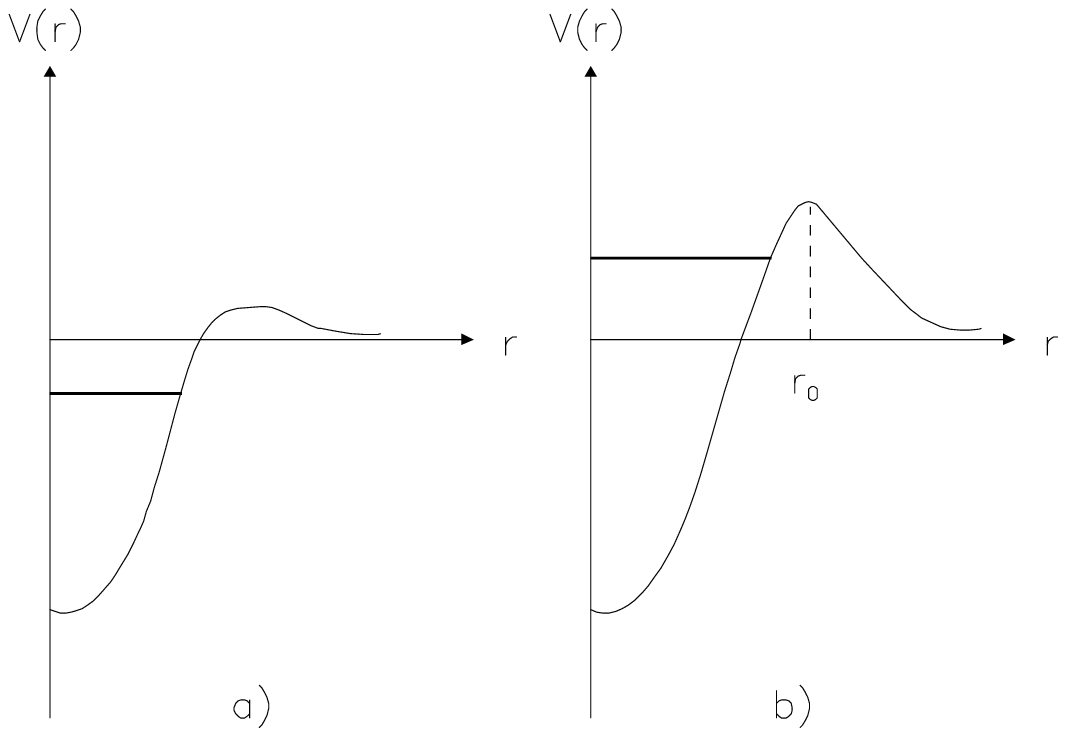,width=14cm}\\
Fig. 4. Stable (a) and non-stable (b) levels in the potential model.
\end{center}
\newpage
\begin{center}
\epsfig{file=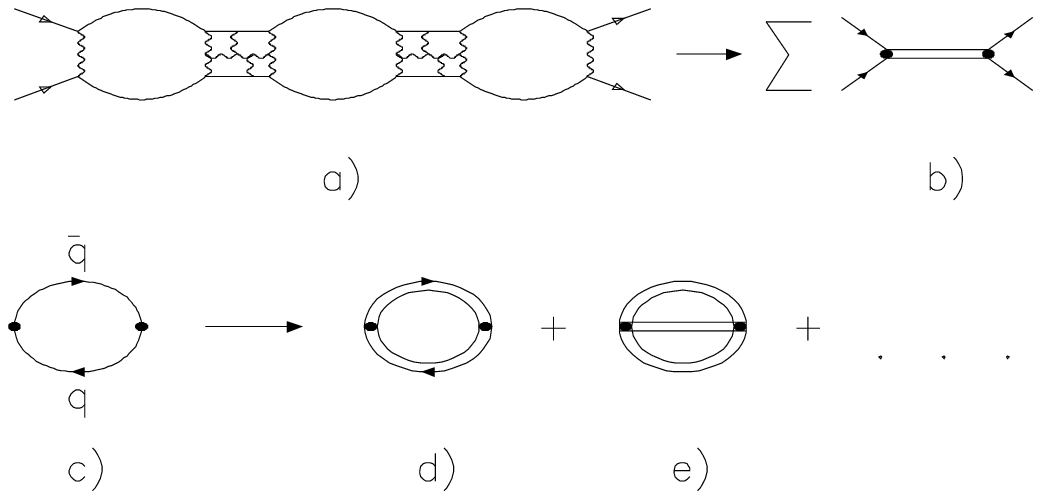,width=14cm}\\
Fig. 5. The $q\bar q$-interaction diagrams at $r<R_{confinement}$
(a) presented as a set of pole terms (b). Saturation of the $q\bar q$-loop
diagram at $r>R_{confinement}$ (c) by a set of meson loop diagrams (d,e).
\end{center}
\newpage
\begin{center}
\epsfig{file=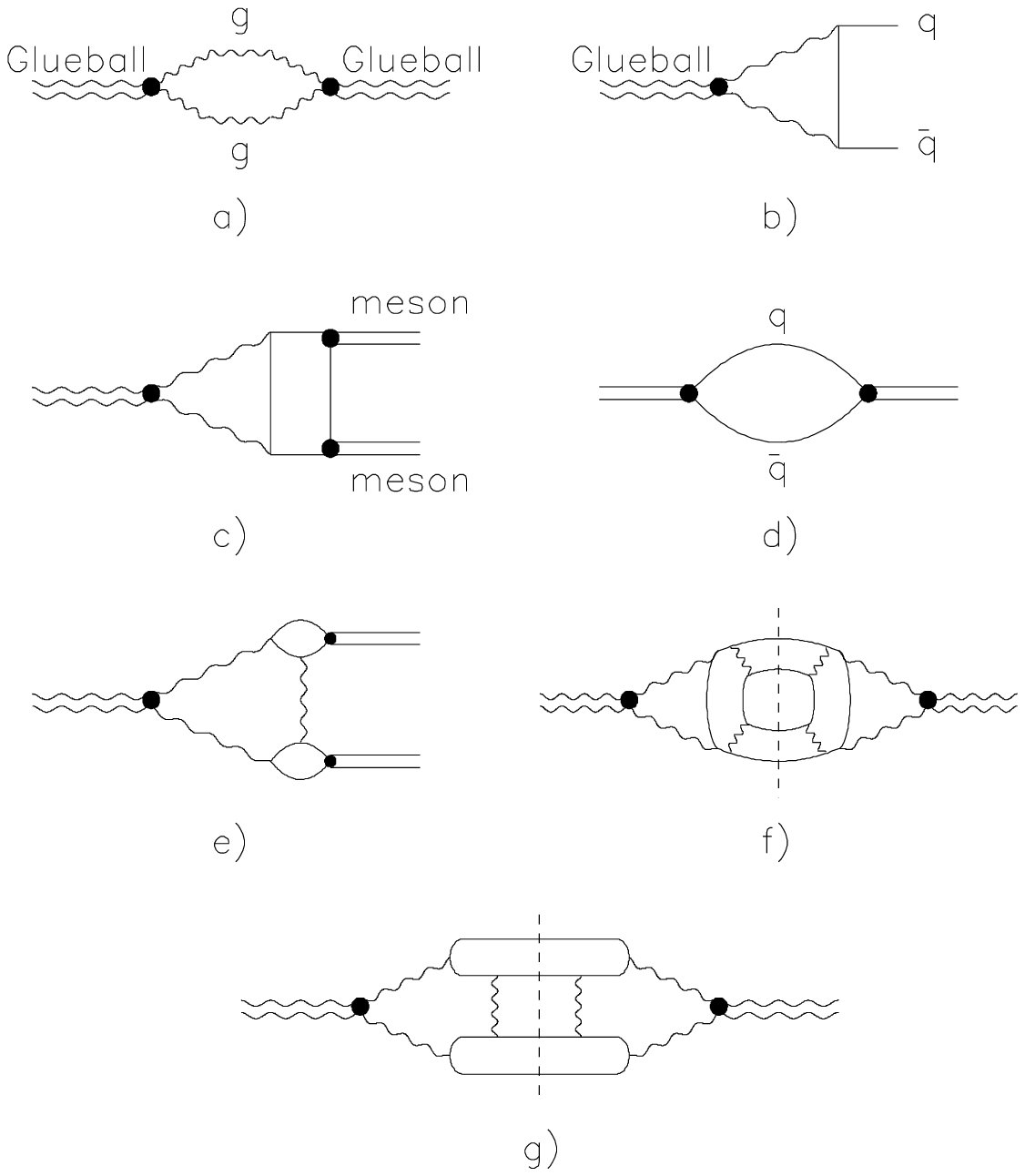,width=14cm}\\
Fig. 6. Diagrams responsible for glueball decay.
\end{center}
\newpage
\begin{center}
\epsfig{file=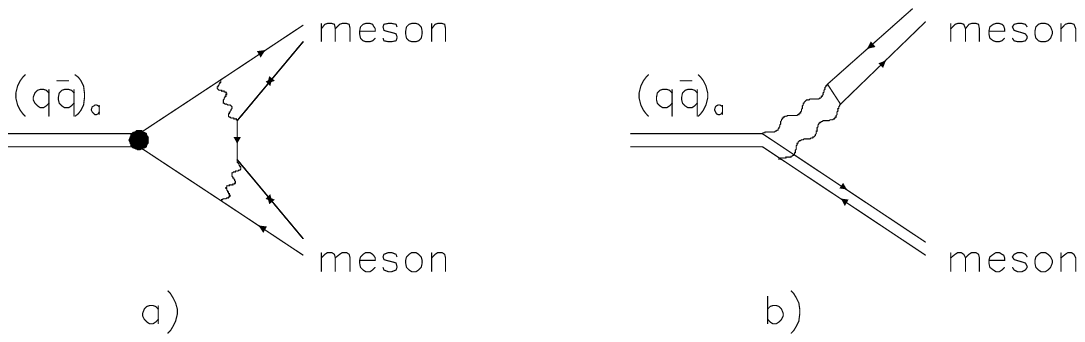,width=14cm}\\
Fig. 7. Diagrams of $(q\bar q)_a$-meson decay.
\end{center}
\newpage
\begin{center}
\epsfig{file=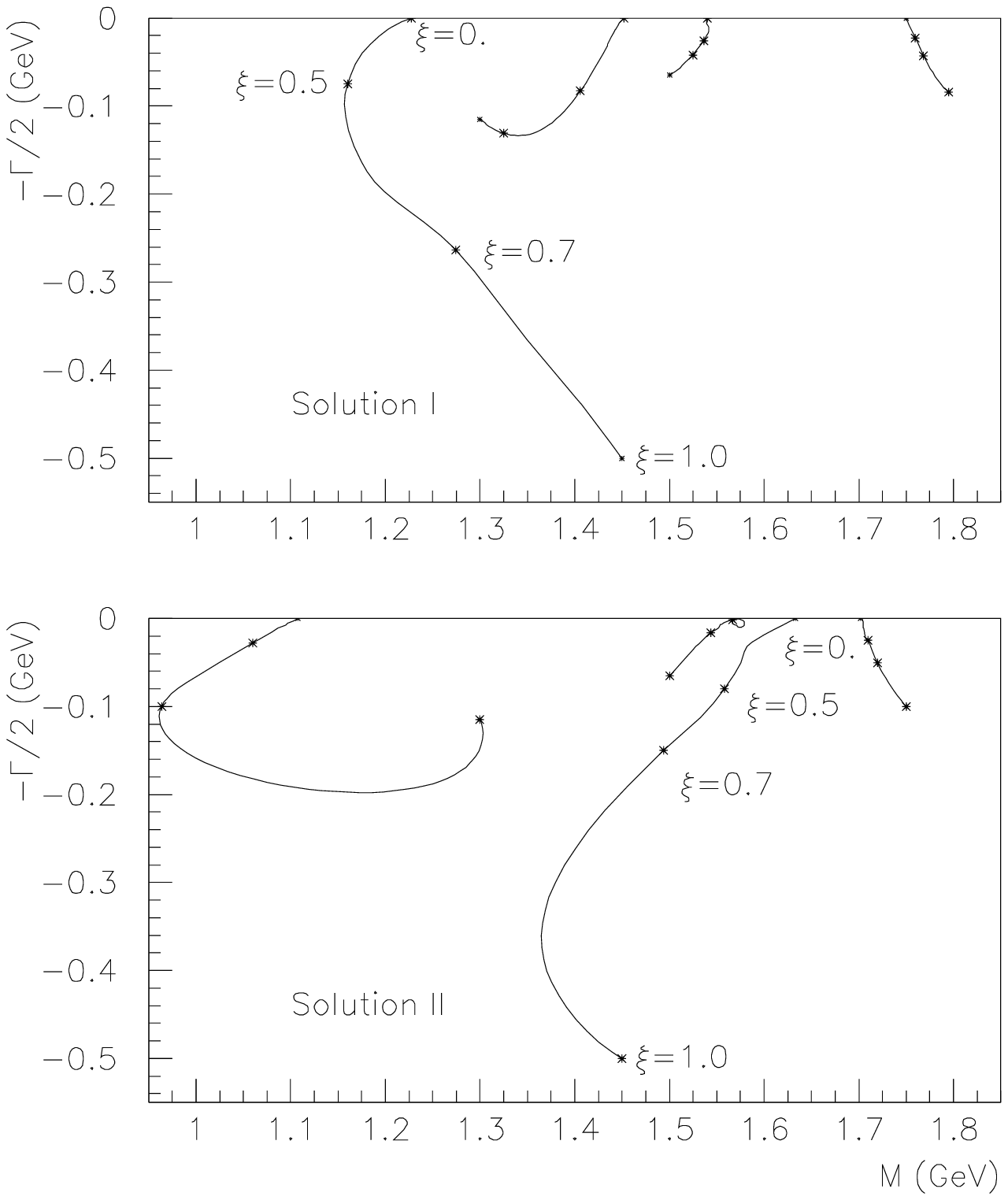,width=14cm}\\
Fig. 8. Complex-$\sqrt{s}$ plane
$(M=Re\sqrt{s},\;\;-\Gamma/2=Im\sqrt{s})$:  location of $00^{++}$
amplitude poles after replacing $g_a \to \xi g_a$.  The case $\xi=0$
gives the positions of masses of input $q\bar q$ states and gluodynamic
glueball; $\xi=1$ corresponds to the real case.
\end{center}

\end{document}